\documentclass[11pt]{article}
\usepackage{graphicx}
\usepackage{subcaption} 
\usepackage{threeparttable} 
\usepackage{tikz} 
\usetikzlibrary{}
\usepackage{float}
\usepackage[left=2cm, right=2cm, top=1.5cm]{geometry}
\usepackage{amsmath}
\usepackage{appendix}
\usepackage{amssymb}
\usepackage{placeins}
\usepackage{nomencl}
\usepackage{authblk}

\newcommand{\parder}[2]{\frac{\partial #1}{\partial #2}}

\newcommand{\dd}[1]{\mathrm{d}#1}

\newcommand{\abs}[1]{\lvert {#1} \rvert}

\DeclareDocumentCommand \dv{ o m m } {
	\IfNoValueTF {#1}{
		\frac{\dd{} #2}{\dd{} #3}
	}{
		\frac{\dd{} #2}{\dd{} #3}|_{\mathrm{#1}}
	}
}

\DeclareDocumentCommand \dvv{ o m m } {
	\IfNoValueTF {#1}{
		\frac{\dd{}^2 #2}{\dd{} {#3}^2}
	}{
		\frac{\dd{}^2 #2}{\dd{} {#3}^2}|_{\mathrm{#1}}
	}
}

\DeclareDocumentCommand \prtdv{ o m m } {
	\IfNoValueTF {#1}{
		\frac{\partial #2}{\partial #3}
	}{
		\frac{\partial #2}{\partial #3}|_{\mathrm{#1}}
	}
}

\DeclareDocumentCommand \prtdvv{ o m m } {
	\IfNoValueTF {#1}{
		\frac{\partial^2 #2}{\partial {#3}^2}
	}{
		\frac{\partial^2 #2}{\partial {#3}^2}|_{\mathrm{#1}}
	}
}

\newcommand{\half}{\frac{1}{2}}

\title{Hierarchical Multiscale Modeling of Positive Corona Discharges}
\author[1]{G. Caliò}
\author[2]{F. Ragazzi}
\author[2]{A. Popoli}
\author[2]{A. Cristofolini}
\author[1]{L. Valdettaro}
\author[1]{C. de Falco}
\author[1]{P. Barbante}
\affil[1]{MOX, Department of Mathematics, Politecnico di Milano, Milano, Italy}
\affil[2]{Department of Electrical, Electronic and Information Engineering, University of Bologna, Bologna, Italy}
\date{}

\usepackage[colorlinks=true,linkcolor=black,anchorcolor=black,citecolor=black,filecolor=black,menucolor=black,runcolor=black,urlcolor=black]{hyperref} 
\usepackage[square, numbers, compress]{natbib} 
\usepackage{fancyhdr} 
\usepackage{multicol}
\makenomenclature

\numberwithin{equation}{section}
\usepackage{afterpage}

\begin{document}
\maketitle


\begin{abstract}
In the field of corona discharges, the complex chemical mechanisms inside the ionization region have prompted the development of simplified models to replicate the macroscopic effects of ion generation, thereby reducing the computational effort, especially in two and three dimensional simulations. We propose a methodology that allows to replace the ionization process with appropriate boundary conditions used by a corona model solving the drift region. We refer to this model as macro-scale, since it does not solve the ionization region. Our approach begins with one dimensional computations in cylindrical coordinates of the whole discharge, where we include a fairly detailed model of the plasma region near the emitter. We refer to this model as full-scale, since all the spatial scales, including the ionization region, are properly taken into account. From these results it is possible to establish boundary conditions for macroscopic simulations. The idea is that, given an emitter radius, the boundary conditions can be used for a variety of geometries that leverage on that emitter as active electrode. 
Our results agree with available experimental data for positive corona discharges in different configurations and with simplified analytical models from literature.
\end{abstract}

\label{ch: Nomenclature}
{
\footnotesize

\renewcommand*{\nompreamble}{\begin{multicols}{2}}
\renewcommand*{\nompostamble}{\end{multicols}}
\setlength{\columnsep}{3em}
\setlength{\nomitemsep}{0.1em}

\begin{thenomenclature} 
\nomgroup{A}
  \item [{$\gamma_j$}]\begingroup Secondary electron emission coefficient 
                                  due to ion impact on the electrodes\nomeqref {0.0}\nompageref{1}
  \item [{$\gamma_{ph}$}]\begingroup Secondary electron efficiency due to photo-ionization\nomeqref {0.0}\nompageref{1}
  \item [{$\mathbf{\Gamma}_k$}]\begingroup Drift-diffusion flux of $k^{th}$ species\nomeqref {0.0}\nompageref{1}
  \item [{$\mathbf{E}$}]\begingroup Electric field\nomeqref {0.0}\nompageref{1}
  \item [{$\mu_k$}]\begingroup mobility of $k^{th}$ species\nomeqref {0.0}\nompageref{1}
  \item [{$\Tilde{S}_k$}]\begingroup Chemical source term of $k^{th}$ species\nomeqref {0.0}\nompageref{1}
  \item [{$\varepsilon$}]\begingroup Dielectric permittivity of the  medium\nomeqref {0.0}\nompageref{1}
  \item [{$\varphi$}]\begingroup Electric potential\nomeqref {0.0}\nompageref{1}
  \item [{$\varphi_{on}$}]\begingroup Onset electric potential\nomeqref {0.0}\nompageref{1}
  \item [{$D_k$}]\begingroup Diffusion coefficient of $k^{th}$ species\nomeqref {0.0}\nompageref{1}
  \item [{$E_{on}$}]\begingroup Onset electric field\nomeqref {0.0}\nompageref{1}
  \item [{$L$}]\begingroup Gap between emitter and collector\nomeqref {0.0}\nompageref{1}
  \item [{$N$}]\begingroup Gas Density\nomeqref {0.0}\nompageref{1}
  \item [{$n_k$}]\begingroup Number density of $k^{th}$ species\nomeqref {0.0}\nompageref{1}
  \item [{$q$}]\begingroup Fundamental charge\nomeqref {0.0}\nompageref{1}
  \item [{$q_k$}]\begingroup Charge of $k^{th}$ species\nomeqref {0.0}\nompageref{1}
  \item [{$R_c$}]\begingroup Collector Radius\nomeqref {0.0}\nompageref{1}
  \item [{$R_e$}]\begingroup Emitter Radius\nomeqref {0.0}\nompageref{1}
  \item [{$r_w$}]\begingroup Wall reflection parameter coefficient \nomeqref {0.0}\nompageref{1}
  \item [{$S_{ph}$}]\begingroup Approximate photo-ionization source term\nomeqref {0.0}\nompageref{1}

\end{thenomenclature}

}

\section{Introduction}

Corona discharges are weakly luminous discharges characterized by spatial non-uniformity, with a distinction between two regions where different physical phenomena occur. The first one is the \textit{plasma region} or \textit{ionization layer}, located near the emitter electrode, where plasma, light generation and a high strength electric field are simultaneously present~\cite{NTP}. The confinement of plasma generation near the emitter electrode is obtained thanks to its geometry; for instance, a thin wire can sustain a strong electric field in its proximity. When the electric field exceeds an onset threshold $E_{on}$, ionization by electron collision begins. Beyond the ionization layer, there is the \textit{drift region}, where the generated ions travel under the influence of the electric field. Further ion generation is usually not possible in this region due to the low electric field~\cite{PlasmaChem}, but ions transfer their momentum to neutral species by collision, resulting in an airflow known as ionic wind. \\
The physical mechanisms discussed above make the full-scale description of the corona phenomenon challenging from a numerical point of view. This is especially true when a gas like air, characterized by complex chemical kinetics, is considered. Historically, the most common approach to the modeling of corona discharges has been fully-coupled monolithic descriptions (i.e.\ where the same mathematical model is applied to the whole physical domain) that relies on simple kinetic models, with chemistry described by Townsend-like models. These models involve electrons and pseudo charged species that represent the total amount of positive and negative ions~\cite{Morrow, Papageorgiou} or only of positive ones~\cite{Montijn,Luque}. Another class of models includes those based on coupled multi-domain simulations, which stem from domain decomposition techniques~\cite{Wettervik} or from asymptotic analysis~\cite{MonPlour, Seimandi}, yet still relying on simplified kinetic schemes. Additionally, there are also models that solve only the drift region and are based on Kaptzov's hypothesis, which states that, once a stable corona regime is established, the electric field on the emitter surface remains constant and equal to the onset value $E_{on}$, which has been experimentally shown to be strongly correlated to the curvature radius of the emitter contact. Based on this assumption, some approximate expressions for charge density and electric field at the edge of the plasma region can be derived and used as boundary conditions (injection law) at the emitter~\cite{adamiak2004simulation, jewell2006design, christen2007simulation, Cagnoni, Coseru, MonPlour}. We can refer to these models as "macro-scale" models, as they do not resolve the ionization events occurring over the small scales of the plasma region. On the opposite, full-scale models also include the plasma region and model the ionizing events with a suitable set of
plasma chemical reactions.\\
Despite their accuracy full-scale models are computationally expensive. Three-dimensional simulations~\cite{Papageorgiou} have been conducted in literature, but considering a simple axisymmetric geometry of the electrodes. Emerging applications such as ionic propulsion~\cite{Belan,Nature} require the simulation of three-dimensional configurations, including the effects of external airflow. To the authors' best knowledge, in these cases only macro-scale models have been used and the plasma region has been replaced by injection laws based on the Kaptzov's hypothesis~\cite{Picella}. From this perspective, macro-scale models appear convenient for three dimensional computations,  as they significantly reduce computational cost by avoiding the coupling between the drift and the plasma region. \\
As one may expect, the predictive accuracy of macro-scale models, mainly depends on the quality of the boundary conditions at the emitter (\emph{injection law} from now on); in Ref.~\cite{Cagnoni} a systematic presentation of different injection laws is made and a simple exponential relation is proposed and shown to be especially convenient for its good numerical performance within coupled simulation algorithms. The exponential injection law presented in Ref.~\cite{Cagnoni} depends on two parameters: in 
addition to the onset field $E_{on}$, another parameter $E_{ref}$ is introduced, that controls the slope of the injection relation near threshold. In the present work we show that the latter feature, which was originally intended only to improve the convergence of nonlinear iteration algorithms, can actually be also exploited in order to more accurately describe the behavior of the emitter near and above onset. Indeed, we show that by fitting the additional parameter to full-scale model simulation, the Kaptzov's hypothesis does not need to be assumed a priori, but it is found a posteriori as a result of full-scale model computations. This is not completely surprising, given the good matching with experimental results of corona models relying on it.\\
Initially, we present results for axisymmetric configurations, where the symmetry property allows the direct usage of one dimensional models. Then our approach is extended to a wire-to-cylinder configuration. Thus in this work a methodological framework is set up for future investigations of more complex arrangements of the electrodes.\\
The one-dimensional simulations in this work are performed using a drift-diffusion-reaction model with a kinetic reaction scheme for dry air that includes six chemical species~\cite{Parent}. The original mechanism of Ref.~\cite{Parent} did not include the effect of photo-ionization. However many literature references point out the role of photo-ionization in providing a seed of electrons that trigger impact ionization and thus sustain the discharge~\cite{MonPlour, dordizadeh, Naidis}. Therefore we present the results both with the original Parent model and with its modification that includes an additional photo-ionization term. \\
The paper is organized as follows. Section~\ref{sec: model} describes the physical models and the underlying assumptions. Section~\ref{sec: NumericalMethods} provides the details about the numerical methods. Section~\ref{sec: Results} shows the validation of the full-scale model against numerical, analytical and experimental results and illustrates the methodology used to obtain the boundary condition for the macro-scale model. Section~\ref{sec: Macroscopic} shows the results of the macroscopic simulations compared with the ones obtained by the full-scale model. Section~\ref{sec: Macroscopic2D} shows macroscopic simulations for truly two-dimensional emitter-collector arrangements and compares them with literature results.

\section{Physical Models}
\label{sec: model}

The main focus of this work is the computation of an injection law at the emitter for a positive corona.\\
In accordance to the phenomenology of positive corona discharges, the ionization length $l_{ion}$ can be considered negligible with respect to the gap $L$ between the electrodes ($l_{ion} \ll L$). The main reason is due to the fact that the high values of the electric field needed to accelerate free electrons and promote ionization reactions are confined in a tiny spatial region that is located around the emitter (active electrode). The overall effect of the plasma region is the injection of positive ions into the drift region. Thus, in the macroscopic model, that has characteristic length $L$, the ionization region can be collapsed over the surface of the emitter and the charge injected into the domain can be accounted for by a suitable boundary condition. If a Kaptzov-like behavior was assumed, the amount of charge on the emitter would be such to maintain the magnitude of the electric field on the emitter constant and equal to the onset value $E_{on}$, computed with the classical Peek's law~\cite{Peek} or with more recent formulas~\cite{Monrolin}. In both cases, for fixed gas conditions, $E_{on}$ is a function only of the radius of the emitter $R_e$. Without directly applying the Kaptzov's hypothesis, but still assuming that the amount of charge entering the drift region is independent from the collector geometry, we aim to determine an injection law of the form $n_p=f(E,R_{e})$, with $n_p$ the number density of the positive ions and $f(E,R_e)$ a nonlinear function of the local electric field and of the emitter radius $R_e$.\\
For a given emitter radius $R_e$ and considering a sufficiently large gap $L$ to avoid the possible insurgence of non local effects, i.e.\ the collector geometry influences $f(E,R_e)$ when the current becomes relatively high~\cite{NegativeCorona}, we can sample the positive ions number density at the end of the ionization region as function of the electric field at the emitter. These number density values are the look-up relation that allows one to set the charge density on the emitter as function of the local electric field. However, in order to enhance the numerical stability and convergence properties of the macro-scale simulation, the raw data can be fitted with a suitable smooth function of $E$. For example, Ref.~\cite{Cagnoni} exploited an exponential law of the form:
\begin{equation}
    n_p(E, R_e)= n_{ref}\exp\left[\frac{E-\Bar{E}_{on}(R_e)}{E_{ref}(R_e)}\right] 
\end{equation}
These strategies require the definition of the ionization length $l_{ion}$ in order to sample the charge density at the boundary between the ionization and drift region. The criterion we used to compute $l_{ion}$ will be explained in Sec.~\ref{sec: IonizationLength}.\\

\subsection{Full-scale Discharge Model}
\label{sec:fullscalemod}

The discharge model is based on the well known drift diffusion approximation. The number density of each plasma component is governed by a continuity equation: 
\begin{align}
    &\frac{\partial n_k}{\partial t} + \nabla \cdot \mathbf{\Gamma}_k = \Tilde{S}_k  \quad k=1,\dots, N_{sp} \label{eq: DD_continuous}
\end{align}
where the flux reads:
\begin{align}
    \mathbf{\Gamma}_k = - D_k \nabla n_k + z_k\mu_k\mathbf{E}\, n_k
\end{align}
$D_k$ and $\mu_k$ are the diffusion and electrical mobility coefficients of the $k^{th}$ species respectively and $z_k=1$ for positive ions, $z_k=-1$ for negative ions and electrons and $z_k=0$ for neutral species. For the neutral species the electrical mobility is set to zero, while for the charged ones the Einstein relation $D_k=\mu_k k_B T_k/q_k$ is used. In this relation $T_k$ is the temperature in Kelvin of the $k^{th}$ species and $k_B$ the Boltzmann constant.\\
The neutral gas is assumed to be Air, approximated as a mixture of $N_2$ and $O_2$ in ratio 3.72 to 1. Ideal gas behavior and atmospheric pressure are assumed.\\
The Local Field Approximation~\cite{KineticsLTP, Alves} is exploited, therefore the electron temperature is assumed to be a function of the local reduced electric field and it is computed with the Boltzmann solver LoKI-B~\cite{LokiB,Tejero-del-Caz_2021}. The ions and neutral gas temperature are both assumed constant and equal to $T=300\,\text{K}$. 
The self-consistent electric field $\mathbf{E}$ is governed, under the quasi-static approximation, by the Poisson equation:
\begin{equation}
       -\nabla \cdot \left(\varepsilon \nabla \varphi\right) = \sum_{k=1}^{N_{sp}}q_k\, n_k \label{eq:poisson}
\end{equation}
where $\varepsilon$ is the medium dielectric permittivity.
Boundary conditions at the electrodes for charged species (ions and electrons) are expressed following the two-stream approach~\cite{Gorin}:
\begin{equation}
    \left\{
    \begin{aligned}
        \mathbf{\Gamma}_i\cdot  \mathbf{n} &= \frac{1-r_w}{1+r_w} n_i\abs{\mathbf{v}^{d,i} \cdot  \mathbf{n}} & i=N_2^+,O_2^+,O_2^-\\
         \mathbf{\Gamma}_e\cdot  \mathbf{n} &= \frac{1-r_w}{1+r_w}n_e\abs{\mathbf{v}^{d,e} \cdot  \mathbf{n}}-\frac{2}{1+r_w}\sum \gamma_j\mathbf{\Gamma}_j\cdot\mathbf{n} & j=N_2^+,O_2^+\\
    \end{aligned}\label{eq: TwoStream}
    \right.
\end{equation}
with $\mathbf{v}^{d,k}=\mu_k\mathbf{E}$ being the drift velocity of ions and electrons $(k=i,e)$. The normal $\mathbf{n}$ at the boundary is directed away from the plasma domain. 
Here $r_w$ is the reflection coefficient and $\gamma_j$ are the secondary electron emission coefficients by ion impact. We have assumed $r_w=0$ and $\gamma_j=10^{-2}$ for all the computations carried out with the full-scale model. \\
Regarding the neutral species, an homogeneous Neumann boundary condition, which corresponds to a non-catalytic wall, has been imposed:
\begin{align}
    \mathbf{\Gamma}_k\cdot  \mathbf{n} &=0 & k=N_2,O_2
    \label{eq:bcneutral}
\end{align}
The boundary conditions for the Poisson equation are Dirichlet boundary conditions. The voltage is imposed at the
emitter and at the collector:
\begin{equation}
\left\{
\begin{aligned}
    \varphi(R_e)&=V_e \\
    \varphi(R_c)&=0
\end{aligned}
\right.
\label{eq:bc_poiss_full}
\end{equation}

The mobilities for each species are taken from Ref.~\cite{Parent} and are reported in Tab.~\ref{tab: Mobilities}, with their original references. 
\begin{table}[H]
    \footnotesize
    \centering
    \begin{threeparttable}
    
     \caption{Ion and electron mobilities in air\textsuperscript{a}}
    \begin{tabular}{ccc}
       Charged Species & Mobility $[\text{\text{m}}^{2}\text{V}^{-1}\text{s}^{-1}]$ & Reference\\ 
        \hline
        \rule{0pt}{1.5em}
        $N^+_2$& $N^{-1}\cdot \min(0.75\cdot10^{23}\cdot T^{-0.5}, 2.03\cdot10^{12}\cdot(E^*)^{-0.5}$&\cite{Sinnott}\\[0.5em]
        $O^+_2$& $N^{-1}\cdot \min(1.18\cdot10^{23}\cdot T^{-0.5}, 3.61\cdot10^{12}\cdot(E^*)^{-0.5}$&\cite{Sinnott}\\[0.5em]
        $O^-_2$&$N^{-1}\cdot \min(0.97\cdot10^{23}\cdot T^{-0.5}, 3.56\cdot10^{12}\cdot(E^*)^{-0.1}$&\cite{Gosho}\\[0.5em]
        $e$& $N^{-1}\cdot 3.74\cdot 10^{19}\cdot \exp(33.5\cdot (\ln(T_e)^{-0.5})$&\cite[Ch. 21]{Grigoriev}\\
    \end{tabular}
    \label{tab: Mobilities}
    \begin{tablenotes}
    \footnotesize
    \item[a] Units: $T_e$ and $T$ are in Kelvin, $N$ is the total number density of the gas in $\text{m}^{-3}$; $E^*$ is the reduced effective electric field ($E^*=|\mathbf{E}|/N$) in units of $\text{V m}^{-2}$.
    \end{tablenotes}
    \end{threeparttable}
\end{table}

The chemical kinetic scheme employed for the computation of the source terms $\Tilde{S_k}$ includes two contributions: chemical reactions and photo-ionization.
Chemical reactions are taken from Ref.~\cite{Parent}: the kinetic model of air includes six species and thirteen reactions and we refer to it as \textit{Parent} model. The original model includes two reactions representing electron beam ionization, that are neglected in the context of corona discharge.
The reactions and their chemical reaction rate constants are reported for convenience in Tab.~\ref{tab: ReactionsPar}.

\begin{table}[H]
    \footnotesize
    \centering
    \begin{threeparttable}
    
     \caption{6 species-13 reactions \textit{Parent} air model\textsuperscript{a}}
    
    \begin{tabular}{clll}
    
        No.& Reaction & Rate Coefficient & Ref. \\
        \hline
        \rule{0pt}{1.5em}
        $1a$& $e^- + N_2 \rightarrow N^+_2 + 2e^-$&$10^{-6}\cdot \exp(-0.0105809 \cdot \ln^2(E^*)- 2.40411 \cdot 10^{-75}\cdot \ln^{46}(E^*))\hspace{0.3em} \text{m}^3/\text{s}$& \cite{Raizer,Mnatsakanyan, Grigoriev}  \\[0.5em]
        $1a$& $e^- + N_2 \rightarrow N^+_2 + 2e^-$&$10^{-6}\cdot \exp(-0.0105809 \cdot \ln^2(E^*)- 2.40411 \cdot 10^{-75}\cdot \ln^{46}(E^*))\hspace{0.3em} \text{m}^3/\text{s}$& \cite{Raizer,Mnatsakanyan, Grigoriev}  \\[0.5em]
        $1a$& $e^- + N_2 \rightarrow N^+_2 + 2e^-$&$10^{-6}\cdot \exp(-0.0105809 \cdot \ln^2(E^*)- 2.40411 \cdot 10^{-75}\cdot \ln^{46}(E^*))\hspace{0.3em} \text{m}^3/\text{s}$& \cite{Raizer,Mnatsakanyan, Grigoriev}  \\[0.5em]
        $1b$& $e^- + O_2 \rightarrow O^+_2 + 2e^-$&$10^{-6}\cdot \exp(-0.0102785 \cdot \ln^2(E^*)- 2.42260 \cdot 10^{-75}\cdot \ln^{46}(E^*))\hspace{0.3em} \text{m}^3/\text{s}$& \cite{Raizer,Mnatsakanyan, Grigoriev} \\[0.5em]
        $2a$& $e^- + O_2^+ \rightarrow O_2 $&$2.0\cdot10^{-13}\cdot(300/T_e)^{0.7}\hspace{0.3em} \text{m}^3/\text{s}$& \cite{Alek}\\[0.5em]
        $2b$& $e^- + N_2^+ \rightarrow N_2$&$2.8\cdot10^{-13}\cdot(300/T_e)^{0.5}\hspace{0.3em} \text{m}^3/\text{s}$& \cite{Kossyi} \\[0.5em]
        $3a$& $O_2^- + N_2^+ \rightarrow N_2 + O_2 $&$2.0\cdot10^{-13}\cdot(300/T)^{0.5}\hspace{0.3em} \text{m}^3/\text{s}$& \cite{Kossyi} \\[0.5em]
        $3b$& $O_2^- + O_2^+ \rightarrow O_2 + O_2 $&$2.0\cdot10^{-13}\cdot(300/T)^{0.5}\hspace{0.3em} \text{m}^3/\text{s}$& \cite{Kossyi} \\[0.5em]
        $4a$& $O_2^- + N_2^+ + N_2\rightarrow O_2 + N_2 + N_2 $&$2.0\cdot10^{-31}\cdot(300/T)^{2.5}\hspace{0.3em} \text{m}^3/\text{s}$& \cite{Kossyi} \\[0.5em]
        $4b$& $O_2^- + O_2^+ + N_2 \rightarrow O_2 + O_2 + N_2$&$2.0\cdot10^{-31}\cdot(300/T)^{2.5}\hspace{0.3em} \text{m}^3/\text{s}$&\cite{Kossyi}  \\[0.5em]
        $4c$& $O_2^- + N_2^+ + O_2\rightarrow O_2 + N_2 + O_2 $&$2.0\cdot10^{-31}\cdot(300/T)^{2.5}\hspace{0.3em} \text{m}^3/\text{s}$&\cite{Kossyi}  \\[0.5em]
        $4d$& $O_2^- + O_2^+ + O_2\rightarrow O_2 + O_2 + O_2$&$2.0\cdot10^{-31}\cdot(300/T)^{2.5}\hspace{0.3em} \text{m}^3/\text{s}$&\cite{Kossyi}  \\[0.5em]
        $5a$& $e^- + O_2 + O_2 \rightarrow O^-_2 + O_2$&$1.4\cdot10^{-41}\cdot(300/T_e)\cdot \exp(-600/T)\cdot \exp\left(\frac{700(T_e-T)}{T_eT}\right)\hspace{0.3em} \text{m}^6/\text{s}$&\cite{Kossyi}  \\[0.5em]
        $5b$& $e^- + O_2 + N_2 \rightarrow O^-_2 + N_2$&$1.07\cdot10^{-43}\cdot(300/T_e)^{2}\cdot \exp(-70/T)\cdot \exp\left(\frac{1500(T_e-T)}{T_eT}\right)\hspace{0.3em} \text{m}^6/\text{s}$& \cite{Kossyi} \\[0.5em]
        $6$& $O^-_2+ O_2 \rightarrow e^- + O_2 + O_2 $&$8.6\cdot10^{-16}\cdot \exp(-6030/T)\cdot \exp(1-\exp(-1570/T)) \hspace{0.3em} \text{m}^3/\text{s}$& \cite[Ch. 2]{Bazelyan} \\[0.5em]
    \end{tabular}
    \label{tab: ReactionsPar}
    \begin{tablenotes}
    \item[a] Units: $T_e$ and $T$ are in Kelvin, $N$ is the total number density of the plasma in $\text{m}^{-3}$; $E^*$ is the reduced effective electric field ($E^*=|\mathbf{E}|/N$) in units of $\text{V m}^{-2}$.
    \end{tablenotes}
    \end{threeparttable}
     
\end{table} 

The functional form of the source term due to chemical reactions is quite standard.
Consider, for example, a set of $M$ reactions involving 
$N_{sp}$ species $k=1\ldots N_{sp}$ and denote by $r_{i,k}$ and $p_{i,k}$ the stoichiometric coefficient of the species $k$ in the reaction $i$, namely
\[
\sum_{k=1}^{N_{sp}} r_{i,k}\ n_k\ \rightleftharpoons{}\ \sum_{k=1}^{N_{sp}} p_{i,k}\ n_k \qquad  i=1\ldots M
\]
Denoting forward reaction rate constant by $c^f_i$, the rate $R^{f}_i$ of the $i$-th forward reaction is 
$$
 R^{f}_i = c^{f}_i \prod_k n_k^{ r_{i,k} }
$$
therefore source term $S_k$ has the form
\begin{equation}
    {S}_k = \sum_{i=1}^M  R^{f}_i\ (p_{i,k}-r_{i,k}) 
\label{eq: SourceTermSk}
\end{equation}

In order to include the effect of photo-ionization, we have followed the approach of Ref.~\cite{MonPlour}. Here the number of photo-ionizing events per unit volume and time at position $\mathbf{R}$ is defined as the integral of the ionization rates times a radiative kernel $g(\mathbf{R},\mathbf{R}')$.
Unlike Ref.~\cite{MonPlour}, where the production of ionized species is modeled with a classical Townsend-like expression, in our model the production of ionized species is the effect of a suitable set of chemical reactions. For air the photo-ionization process is usually triggered by photons emitted by excited molecular nitrogen $N_2$. These photons, in their turn, ionize oxygen molecules via the reaction $O_2+h\nu \rightarrow O_2^+ + e$~\cite{dordizadeh}. $N_2$ excitation is usually due to energetic impact with
electrons (i.e.\ electrons transfer energy to $N_2$): it is therefore reasonable to assume that the photo-ionization rate depends on the rate of reaction $1a$ reported in Table \ref{tab: ReactionsPar}. 
The photo-ionization source term therefore reads:
\begin{equation}
S_{ph}(\mathbf{R}) =  \int  \gamma_{ph}S^{1a}(\mathbf{R'})\frac{g(\mathbf{R}, \mathbf{R}')}{4\pi|\mathbf{R}-\mathbf{R}'|^2}  \, d^3\mathbf{R}'
\label{eq:photion3D}
\end{equation}
where $S^{1a}$ is the rate of reaction $1a$ and $\gamma_{ph}$ is the secondary electron efficiency. $\gamma_{ph}$ takes into account the fact that only a fraction of the energetic collisions between $N_2$ and electrons leads to an excited state. The expression of $\gamma_{ph}$ is taken from Ref.~\cite{beni19}:
\begin{equation}
    \gamma_{ph}=\left(0.03+\frac{15.7\cdot10^{-21}}{\abs{\mathbf{E}}/N}\right)\frac{p_q}{p_q+p}
    \label{eq:gamma_ph}
\end{equation}
where $p_q = 30\, \text{Torr}$ is a quenching pressure and $p$ is the gas pressure (atmospheric one in our case).
The expression of the three dimensional photo-ionization kernel $g(\mathbf{R},\mathbf{R}')$ is taken from~\cite{Bourdon, Zheleznyak}:
\begin{equation}
    g(\mathbf{R},\mathbf{R'})= \frac{\exp({-\lambda_1 |\mathbf{R}-\mathbf{R'}|})-\exp({-\lambda_2|\mathbf{R}-\mathbf{R'}|})}{\ln({\lambda_2 /\lambda_1})|\mathbf{R}-\mathbf{R'}|}
    \label{eq:photokernel}
\end{equation}
where $\lambda_1 = \chi_{\text{min}} \, p_{O_2}$, $\lambda_2 = \chi_{\text{max}} \, p_{O_2}$, and $\chi_{\text{min}} = 0.035 \, \text{Torr}^{-1} \, \text{cm}^{-1}$, $\chi_{\text{max}} = 2 \, \text{Torr}^{-1} \, \text{cm}^{-1}$ and where $p_{O_2}$ is the partial pressure of molecular oxygen. 
Considering atmospheric pressure, $p_{O_2} = 150 \, \text{Torr}$, thus $1/\lambda_1 = 1905 \, \mu\text{m}$ and $1/\lambda_2 = 33.3 \, \mu\text{m}$ \cite{MonPlour}.\\
The geometry of the full-scale computation is cylindrical (see Fig.~\ref{fig:geometry}) and the problem is invariant along the longitudinal axis $\mathbf{e}_z$. It is convenient to decompose the position vector $\mathbf{R}$ into a vector $\mathbf{r}$ lying in the reference plane and into a vector $z\mathbf{e}_z$ along the longitudinal axis: $\mathbf{R}= \mathbf{r} + z\mathbf{e}_z$. The source term $S^{1a}$ and $\gamma_{ph}$ do not depend on $z$ and the general three-dimensional kernel of Eq.~\ref{eq:photokernel} can be integrated along the longitudinal axis $\mathbf{e}_z$. This integration defines a new two-dimensional photo-ionization kernel:
\begin{equation}G(\mathbf{r},\mathbf{r}')\equiv G(\rho\equiv|\mathbf{r} - \mathbf{r}'|) = \int  \frac{g(\mathbf{R},\mathbf{R'})}{4 \pi |\mathbf{R}-\mathbf{R'}|^2} \, dz'
\end{equation}
The integration along $z$ provides an explicit expression of the two-dimensional kernel $G(\mathbf{r},\mathbf{r}')$ \cite{MonPlour}:
\begin{equation}
    G(\rho) = \frac{\rho}{4 \pi \ln \left( \frac{\lambda_2}{\lambda_1} \right)} \left[ \lambda_1^3 \mathcal{G}_{3,0}^{0,3} \left(\bigg|\begin{array}{c} 0 \\  \, \left( -\frac{1}{2}, -1, -\frac{3}{2} \right)\end{array} \, \bigg| \, \left(\frac{\lambda_1 \rho}{2}\right)^2 \right) - \lambda_2^3 \mathcal{G}_{3,0}^{0,3} \left(\bigg|\begin{array}{c} 0 \\  \, \left( -\frac{1}{2}, -1, -\frac{3}{2} \right) \end{array}\, \bigg| \, \left(\frac{\lambda_2 \rho}{2}\right)^2 \right) \right]
\end{equation}
where $\mathcal{G}\!{\footnotesize \begin{array}{c} m\ n\\p \ q\end{array}}\!
\left(\begin{array}{c}
    a_1,\dots, a_p\\
    b_1, \dots, b_q
\end{array}\! \bigg| \, z\right)$ is the Meijer-G function \cite{abramowitz1968handbook}.\\
Therefore the photo-ionization source term of Eq.~\ref{eq:photion3D} now becomes:
\begin{equation}
    S_{ph}(\mathbf{r}) = \int_{\Omega} \gamma_{ph} S^{1a}(\mathbf{r}') G(\mathbf{r},\mathbf{r}')\, d^2\mathbf{r}'
    \label{eq:photoion2D}
\end{equation}
The function $G(\mathbf{r},\mathbf{r'})$ strongly decreases as a function of its argument $\rho$ and it is convenient to expand it inside the integral~\cite{MonPlour}:
\begin{equation}
    S_{ph}(\mathbf{r}) = G(|\mathbf{r}|) \int_{\Omega}\gamma_{ph} S^{1a}(\mathbf{r}') d^2\mathbf{r}' + \nabla G(|\mathbf{r}|) \cdot \int_{\Omega} \gamma_{ph} S^{1a}(\mathbf{r}') \mathbf{r'} d^2\mathbf{r}' + O(\rho^2)
\end{equation}
One can define the leading order term $S_{ph,0}(\mathbf{r})= G(|\mathbf{r}|) \int_{\Omega} \gamma_{ph}S^{1a}(\mathbf{r}') d^2\mathbf{r}'$ and the first order term $S_{ph,1}(\mathbf{r})=\nabla G(|\mathbf{r}|) \cdot \int_{\Omega}\gamma_{ph} S^{1a}(\mathbf{r}')\mathbf{r}' d^2\mathbf{r}'$ of the expansion:
\begin{equation}
    S_{ph}(\mathbf{r}) = S_{ph,0}(\mathbf{r}) + S_{ph,1}(\mathbf{r}) + O(\rho^2)
\end{equation}
The leading order term $S_{ph,0}(\mathbf{r})$ is the dominant one and the photo-ionization source term reduces to:
\begin{equation}
    S_{ph}(\mathbf{r}) \approx S_{ph,0}(\mathbf{r})
    \label{eq: photion}
\end{equation}
Finally the total source term $\Tilde{S}_k$ is:
\begin{equation}
    \left\{
    \begin{aligned}
        &\Tilde{S_e}        = S_e +S_{ph}\\
        &\Tilde{S}_{O_2} = S_{O_2}  - S_{ph}\\
        &\Tilde{S}_{O_2^+}=S_{O_2^+}  + S_{ph}\\
        &\Tilde{S}_j  =S_j  \qquad j=N_2,N_2^+,O_2^-
    \end{aligned}
    \right.
\end{equation}

\subsection{Macroscopic model of the Drift region}
\label{sec:macromodel}

In the drift region the reaction terms are negligible because of the low magnitude of the electric field, as confirmed by the asymptotic analysis of Ref.~\cite{MonPlour}.
Therefore, at stationary state, the drift-diffusion model in this portion of the domain reduces to:
\begin{equation}
   \begin{cases}
       \nabla \cdot \boldsymbol{\Gamma}_p = 0 \label{eq: DD_macro}\\
       \boldsymbol{\Gamma}_p = - D_p \nabla n_p + \mu_p \mathbf{E}\,n_p
   \end{cases}
\end{equation}
where $n_p$, $\mu_p$ and $D_p$ are respectively the number density, the mobility and diffusion coefficient of the charge carriers, namely the positive ions. In practice we replace the two positive ions of the full-scale model, $N_2^+$ and $O_2^+$, with a positive overall pseudo species.\\
In the electric quasi-static approximation, the electric field $\mathbf{E}$ is equal to the gradient of its potential $\varphi$. Hence the Poisson equation reads as: 
\begin{equation}
    -\nabla \cdot \left(\varepsilon \nabla \varphi\right) = q n_p\label{eq: Poisson_macro}
\end{equation}
The Poisson equation is completed with Dirichlet boundary conditions:
\begin{equation}
\left\{
\begin{aligned}
    \varphi(R_e)&=V_e \\
    \varphi(R_c)&=0
\end{aligned}
\right.
\label{eq: bc_poisson}
\end{equation}
that, we notice, are equal to the ones of Eq.~\ref{eq:bc_poiss_full}.
As explained in Sec.~\ref{sec: model}, the continuity equation is completed with the following boundary conditions:
\begin{align}
    n_p(R_e)&=\max(f(E,{R_e}),n_{min})\label{eq: bc_emi_macro} \\
    n_p(R_c)&=n_{min}\label{eq: nmin}
\end{align}
Eq.~\ref{eq: bc_emi_macro} is the injection law based on the results of the full-scale simulations and will be discussed in Sec.~\ref{sec: LookUpTable}.
$n_{min}$ is a constant value representing a minimum level of charged species number density, that we set equal to $10^9\, \text{m}^{-3}$ in all the simulations. The effect of the choice of the value of $n_{min}$ on the results is assessed in Appendix~\ref{sec:appa}.\\
In the macroscopic model the mobility is assumed to be constant in contrast to the full-scale model. This assumption is common in other literature contributions~\cite{Coseru,MonPlour,Picella}. For all the simulations we assume $\mu = 2\cdot10^{-4}\, \text{m}^2\, \text{V}^{-1}\,\text{s}^{-1}$. This is the same value adopted by Ref.~\cite{Coseru}. Moreover it is approximately the limiting value for positive air ions in the \textit{Parent} model assuming a gas temperature of $T=300\, \text{K}$.

\section{Numerical Method}
\label{sec: NumericalMethods}

In this section we describe the numerical methods used to solve the equations stemming from the models 
described above. We start in Sec.~\ref{sec:num_fullscale} by considering the full-scale model of Sec.~\ref{sec:fullscalemod} as it is the most complex to deal with. Indeed since the method required to solve the macroscopic model is just a special simplified case of that of Sec.~\ref{sec:num_fullscale}, we briefly describe such simplifications and modifications in Sec.~\ref{sec:num_macromodel}.

\subsection{Full-scale model}
\label{sec:num_fullscale}

\begin{figure}
    \centering
    \includegraphics[width=0.3\textwidth]{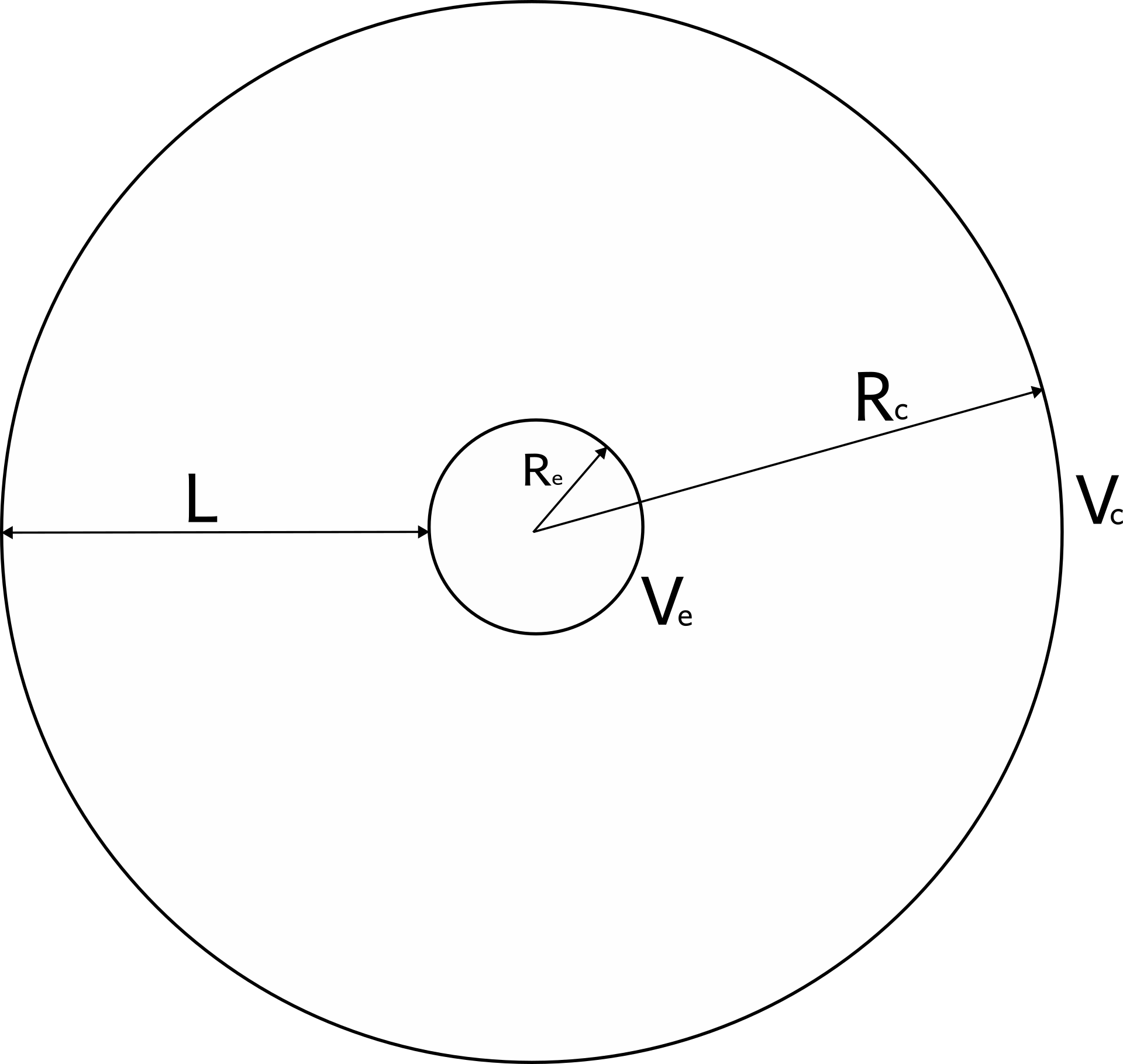}
    \caption{Wire-in-Cylinder configuration of the electrodes. Emitter electrode is magnified for sake of clarity}
    \label{fig:geometry}
\end{figure}
The full-scale model is applied to a configuration with concentric cylindrical electrodes (see Fig.~\ref{fig:geometry}).
The spatial discretization of the drift-diffusion equations \ref{eq: DD_continuous} and of Poisson equation \ref{eq:poisson} is based on a finite volume cell centred scheme.
Due to the symmetry of the geometry, we work in cylindrical coordinates and take into account only the dependence on the radial coordinate $r$.\\
A sketch of the computational mesh is shown in Fig.~\ref{fig:mesh}.
The domain is divided into $N_C$ interior cells plus two ghost cells. The unknowns (species number densities and potential, we use the generic symbol $n_i$ in Fig.~\ref{fig:mesh}) are located at cell centres, except for $n_1$ and $n_{N_{C+2}}$, that are located on the physical boundaries of the domain (i.e.\ the surface of the electrodes). Continuous lines are the physical boundaries of the domain, dashed lines are the interfaces between interior cells and dotted lines are the internal interfaces of the two ghost cells (the other two being the physical interfaces).
\begin{figure}[ht]
	\centering
	{
	\begin{tikzpicture}
        \pgfmathsetmacro{\angle}{40}
        
		\pgfmathsetmacro{\d}{1}
		\pgfmathsetmacro{\s}{0.3}
		\pgfmathsetmacro{\xnl}{-3 *\d}
		\pgfmathsetmacro{\xnc}{0}
		\pgfmathsetmacro{\xnr}{4*\d}
		\pgfmathsetmacro{\xir}{0.5*\xnr + 0.5*\xnc}
		\pgfmathsetmacro{\xil}{0.5*\xnl + 0.5*\xnc}
        \pgfmathsetmacro{\xnll}{-5*\d}
        \pgfmathsetmacro{\xnrr}{7*\d}
        \pgfmathsetmacro{\xilll}{-6*\d} 
        \pgfmathsetmacro{\xill}{0.5*\xnll + 0.5*\xnl}
		\pgfmathsetmacro{\xirrr}{8.5*\d}
        \pgfmathsetmacro{\xirr}{0.5*\xnrr + 0.5*\xnr}

        \pgfmathsetmacro{\radius}{0.5*\d/sin(\angle)} 
        \pgfmathsetmacro{\pad}{\radius*(1-cos(\angle))} 
        
		\filldraw [black] (0,0) circle (2pt);
		\filldraw [black] (\xnr,0) circle (2pt);
		\filldraw [black] (\xnl,0) circle (2pt);
		\node() at (0,2.5*\s) {$n_i$};
		\node() at (\xnr,2.5*\s) {$n_{i+1}$};
		\node() at (\xnl,2.5*\s) {$n_{i-1}$};
        \draw[dashed] (\xil-\pad,-\d/2) arc[start angle=-\angle, end angle=\angle, radius=\radius];
        \draw[dashed] (\xir-\pad,-\d/2) arc[start angle=-\angle, end angle=\angle, radius=\radius];
		\node() at (\xir,\d+\s) {$i+\half$};
		\node() at (\xil,\d+\s) {$i-\half$};
		
		\draw [<->] (\xil,-\d-0.1) -- (\xir,-\d-0.1) node[midway,above] {${\Delta r}_i$};

         \filldraw [black] (\xilll,0) circle (2pt) node() at (\xilll,2.5*\s) {$n_1$};
        \filldraw [black] (\xnll,0) circle (2pt) node() at (\xnll,2.5*\s) {$n_2$};
        \filldraw [black] (\xnrr,0) circle (2pt) node() at (\xnrr,2.5*\s) {$n_{N_{C+1}}$};
        \filldraw [black] (\xirrr,0) circle (2pt) node() at (\xirrr,2.5*\s) {$n_{N_{C+2}}$};
        \draw (\xilll-\pad,-\d/2) arc[start angle=-\angle, end angle=\angle, radius=\radius];
        \draw[dashed] (\xill-\pad,-\d/2) arc[start angle=-\angle, end angle=\angle, radius=\radius];
        \draw[dashed] (\xirr-\pad,-\d/2) arc[start angle=-\angle, end angle=\angle, radius=\radius];
        \draw (\xirrr-\pad,-\d/2) arc[start angle=-\angle, end angle=\angle, radius=\radius];
        \draw[dotted] (\xnll-\pad,-\d/2) arc[start angle=-\angle, end angle=\angle, radius=\radius];
        \draw[dotted] (\xnrr-\pad,-\d/2) arc[start angle=-\angle, end angle=\angle, radius=\radius];
	\end{tikzpicture}
	}
    \caption{Sketch of the computational mesh.}
    \label{fig:mesh}
\end{figure}
For sake of clarity we rewrite Eqs.~\ref{eq: DD_continuous} and \ref{eq:poisson} in the new cylindrical reference system. The continuity drift-diffusion equation reads:
\begin{equation}
      \frac{\partial n_k}{\partial t} + \frac{1}{r}\parder{\left(r\Gamma_k\right)}{r}  = \Tilde{S}_k
      \label{eq:DD_cyl}
\end{equation}
where $\Gamma_k=-D_k\parder{n_k}{r}+z_k \mu_k E_r\, n_k$ is the flux and $E_r=-\parder{\varphi}{r}$ the component of the electric field along the $r$ axis.
The Poisson equation for the electric field now reads:
\begin{equation}
       \frac{1}{r} \parder{}{r}\left(r\parder{\varphi}{r}\right) = -\frac{1}{\varepsilon}\sum_{k=1}^{N_{sp}}q_k n_k 
       \label{eq:poisson_cyl}
\end{equation}
In order to discretize Eqs.~\ref{eq:DD_cyl} and \ref{eq:poisson_cyl} let's multiply each side of the equations by $r$ and then integrate over each cell. The standard approximations used in the finite volume method lead to the following equations for species density $n_k$ and for electric potential $\varphi$, written for the $i^{th}$ interior cell:
\begin{equation}
    \left.\frac{\partial n_k}{\partial t}\right|_i+
    \frac{1}{r_i{\Delta r}_i}\left[\left(r\Gamma_k\right)|_{i+1/2}-\left(r\Gamma_k\right)|_{i-1/2}\right]=\Tilde{S}_{k}|_i
   \label{eq:DD_integrated_compact}
\end{equation}
\begin{equation}
    \frac{1}{r_i{\Delta r}_i}\left[\left.\left(r\parder{\varphi}{r}\right)\right|_{i+1/2}-\left.\left(r\parder{\varphi}{r}\right)\right|_{i-1/2}\right]=-\frac{1}{\varepsilon}\sum_{k=1}^{N_{sp}}(q_k n_k)|_i
   \label{eq:poisson_integrated}
\end{equation}
where we denote with the halved labels the cell interfaces and with the integer ones the cell centre.
$\left(r\Gamma_k\right)|_{i+1/2}$ and $\left.\left(r\parder{\varphi}{r}\right)\right|_{i+1/2}$ are numerical fluxes and we need a suitable expression for them.
The numerical flux for Eq.~\ref{eq:DD_integrated_compact} is computed with the exponential~\cite{allen55}, or Scharfetter-Gummel~\cite{schar69}, scheme:
\begin{equation}
\left(r\Gamma_k\right)|_{i+1/2}=\frac{D_k|_{i+1/2}}{\ln({r_{1+1}/r_i})}\left(B(X_k|_{i+1/2})n_k|_i-B(-X_k|_{1+1/2})n_k|_{i+1}\right)\label{eq: numerical_flux_continuity}
\end{equation}
where $B(X)=X/(e^X-1)$ is Bernoulli's function and $X_k|_{i+1/2}=z_k\mu_k|_{i+1/2}(\varphi_{i+1}-\varphi_i)/D_k|_{i+1/2}$.
The numerical flux for Eq.~\ref{eq:poisson_integrated} is:
\begin{equation}
\left.\left(r\parder{\varphi}{r}\right)\right|_{i+1/2}=\frac{\varphi_{i+1}-\varphi_i}{\ln(r_{i+1}/r_i)}\label{eq: numerical_flux_poisson}
\end{equation}
The boundary conditions are applied in a straightforward way. For the drift-diffusion equations the numerical flux at the physical boundaries of the domain
(continuous lines in Fig.~\ref{fig:mesh}) is computed with Eqs.~\ref{eq: TwoStream} for the ionized species and with Eq.~\ref{eq:bcneutral} for neutral ones, instead of Eq.~\ref{eq: numerical_flux_continuity}. For the Poisson equation we impose $\varphi_1=V_e$
and $\varphi_{N_{C+2}}=0$.\\
The problem at hand is characterized by strong gradients of the electric field and of the species number densities near the emitter electrode. Therefore, the mesh is built following a non uniform distribution along the radial direction. The length $\Delta r_i$ of the cells follows an hyperbolic tangent distribution. The advantage is the possibility to cluster a higher number of cells in the vicinity of the emitter electrode. Moreover the distribution has been made asymmetric, in order to increase the refinement near the emitter. \\
The time integration of the system \ref{eq:DD_integrated_compact} is performed with the SUNDIALS~\cite{sundials} module available inside the MATLAB software. We have selected the CVODE stiff solver, that implicitly integrates the system of equations with a variable order method. In our case we have chosen a third order scheme. \\

\subsection{Computation of $S_{ph}$ in cylindrical coordinates}
We express the integral appearing in Eq.~\ref{eq: photion} in cylindrical coordinates:
\begin{equation}
    S_{ph}(r,\theta) =G(r) \int_{\Omega}\gamma_{ph}  S^{1a}(r') \, r'dr'd\theta' 
\end{equation}
Given the symmetry of the problem the argument of the integral does not depend on the variable $\theta'$ and defining $\gamma_{eff}=2\pi\gamma_{ph}$ it follows:
\begin{equation}
    S_{ph}(r) = G(r) \int_{R_c}^{R_e} \gamma_{eff} S^{1a}(r') \, r'dr' 
    \label{eq:sph_cyl}
\end{equation}
The integral appearing in Eq.~\ref{eq:sph_cyl} is computed with the standard mid-point rule of finite volume method.

\subsection{Macroscopic model}
\label{sec:num_macromodel}

The governing equations of the macroscopic model are also discretized with a finite volume method. In particular, the numerical flux of the discrete version of Eq.~\ref{eq: DD_macro} is expressed by Eq.~\ref{eq: numerical_flux_continuity}, while for Eq.~\ref{eq: Poisson_macro} we resort to Eq.~\ref{eq: numerical_flux_poisson}.
The non linear system of equations obtained from the discretization procedure is solved, together with the boundary conditions given by Eqs.~\ref{eq: bc_poisson}, \ref{eq: bc_emi_macro} and~\ref{eq: nmin},  via the \textit{fsolve} function of MATLAB.

\section{Full-scale model results}
\label{sec: Results}

The simulations are performed imposing on the emitter a voltage that varies in a quasi-stationary fashion:
\begin{equation}
    V_e(t)=V_0+ 
    \left\{
    \begin{aligned}
    &0 &\qquad0\leq t<1\\
    &V_f \frac{t-1}{T} & 1\leq t<T-1\\
    &V_f &\qquad T-1\leq t \leq 2T\\
    \end{aligned}
    \right.
\end{equation}
With a suitable choice of $T$ the voltage varies so slowly in time, that at each instant the solution is practically stationary. This procedure allows to retrieve the entire output dataset over the chosen range of voltage $[V_0,V_f]$ in a single simulation. In our simulations we choose $T=200 \, \text{s}$.\\
The current produced by the corona discharge is computed according to the Sato's formula~\cite{Sato,Ragazzi202412545}:
\begin{align}
    I=\frac{1}{V_e}\iiint_{\Omega}\left(\sum_{k=1}^{N_{sp}}{q_k\mathbf{\Gamma_k}}\right)\cdot\mathbf{E}_s d\mathbf{V} \quad [\text{A}]
\end{align}
that, in a one-dimensional geometry in cylindrical coordinates, reduces to:
\begin{align}
    J=\frac{2\pi}{V_e}\int_{R_e}^{R_c}  \left(\sum_{k=1}^{N_{sp}} q_k\Gamma_k\right) {E}_s r dr \quad[\text{A/m}]
\end{align}
where $E_s$ is the static or Laplacian electric field, obtained considering only the applied voltage and neglecting the charge density inside the volume.

\subsection{Validation}
\label{sec:validation}

\begin{figure}[h]
    \centering

        \includegraphics[width=0.55\linewidth]{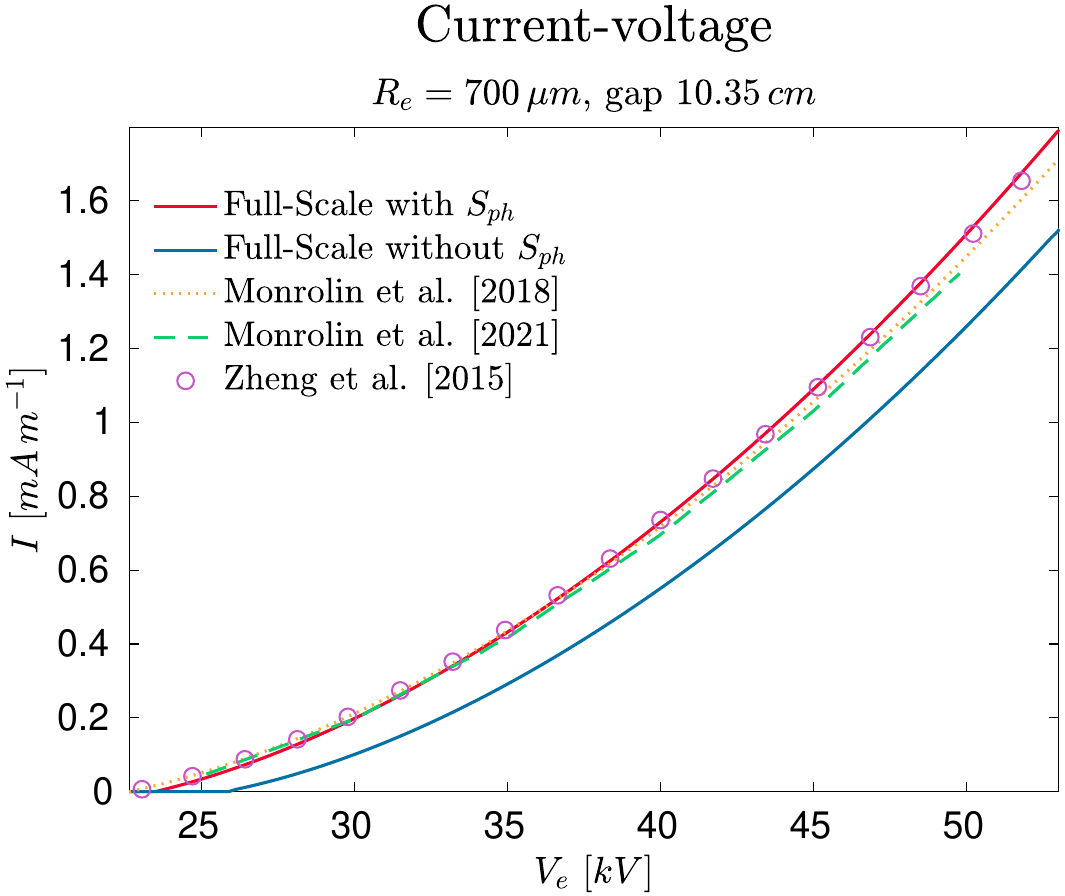} %
    \caption{Current-voltage curves obtained from experimental results (Zheng et al.\ 2015) \cite{Zheng}, numerical two-scale model (Monrolin et al.\ 2021) \cite{MonPlour}, analytical model (Monrolin et al.\ 2018) \cite{Monrolin} and 1-D simulation with Parent model with and without photo-ionization. Emitter radius $700\,\mu\text{m}$, gap $10.35\, \text{cm}$.}
    \label{fig:iv_curve_validation}
\end{figure}

The results of the full-scale model are compared with the experimental measurements of Ref.~\cite{Zheng} in order to provide a validation of our assumptions. In Fig.~\ref{fig:iv_curve_validation} the experimental current-voltage data are confronted with the numerical data obtained with the plasma model presented in Sec.~\ref{sec:fullscalemod}. We performed the computations both with the original Parent kinetic model without photo-ionization and with our modified version that includes photo-ionization.

Our full-scale model with the inclusion of photo-ionization appears to be in good agreement with the experimental data.
The emitter radius $R_e$ is quite large, being equal to $700\,\mu\text{m}$; the local electric field, for purely geometric reasons, is not too high and the contribution of photo-ionization is not negligible.
The small difference in the slope of the curve with respect to Refs.~\cite{Monrolin} and~\cite{MonPlour} is due to the mobility of ions which they take constant, while in our model it is a function of the electric field and of temperature.
The current is well predicted for almost all applied voltages, excluding under $V_e=25\, \text{kV}$, a value that is however close to the onset voltage, estimated to be $V_{on}\approx 22.1\,\text{kV}$ in Ref.~\cite{Zheng}. The addition of other chemical species could improve the quality of predictions in the neighborhood of the onset voltage.
However, we should point out that in this contribution we aim at devising an injection law to be used after the onset and therefore a very precise estimation of the onset voltage is not of fundamental importance.
Onset conditions are the limit above which current is generated in the corona~\cite{Monrolin,Zheng}. The onset, or ignition, voltage $V_{on}$ can be accurately computed as the solution of a suitable eigenvalue problem~\cite{benilov_onset}. Instead of solving the eigenvalue problem, we have opted to identify as onset the voltage at which the current starts to be greater than zero. Because of the limitations of our model and discretization errors of numerical approximations, it is reasonable to consider the onset of the discharge when the computed current reaches a threshold value set equal to $10^{-6}\, \text{A/m}$.
In Table~\ref{tab: OnsetErrors} the onset electric field estimated from our simulations for three different emitter radii and a gap $L=9\,\text{cm}$, considering the model with and without photo-ionization, is compared with the one predicted according to Ref.~\cite{Monrolin}.
The numerically computed onset electric field $E_{on}$ with photo-ionization is in good agreement with the analytical formula of Ref.~\cite{Monrolin}. This shows that the choice of a threshold current of $10^{-6}\, \text{A/m}$ to identify the discharge onset is a reasonable criteria. The onset electric field without photo-ionization is instead systematically higher, because a stronger applied voltage is needed to initiate the discharge.
\begin{table}[H]
    \footnotesize
    \centering
     \caption{Comparison of the estimated onset field with respect to theoretical predictions of Ref.~\cite{Monrolin}. Here the onset criterion is based on a threshold current value of $10^{-6}\ \text{A/m}$.}
    \begin{tabular}{cccc}
       Emitter Radius $[\mu\text{m}]$ & $E_{on} [\text{V/m}]$~\cite{Monrolin} & $E_{on} [\text{V/m}]$ [No $S_{ph}$]& $E_{on} [\text{V/m}]$ [With $S_{ph}$] \\ 
        \hline
        \rule{0pt}{1.5em}
        $50$&  $1.479\;10^7$&$ 1.975\;10^7$& $1.479\;10^7$\\[0.5em]
        $300$& $8.044\;10^6$& $9.472\;10^6$& $ 8.244\;10^6$\\[0.5em]
        $700$&  $6.471\;10^6$&$7.352\;10^6$& $6.690\;10^6$\\[0.5em]
    \end{tabular}
    \label{tab: OnsetErrors}
 
\end{table}
The computation without photo-ionization underestimates the current for the same applied voltage. 
The absence of the photo-ionization source term $S_{ph}$ is clearly visible on the current-voltage curve: the discharge produces the same current with a higher applied voltage on the emitter. The reason could be found in the lack of sufficiently energized electrons inside the plasma region that can sustain the discharge at lower voltages.

\begin{figure}[H]
        \centering
        \includegraphics[width=0.6\linewidth]{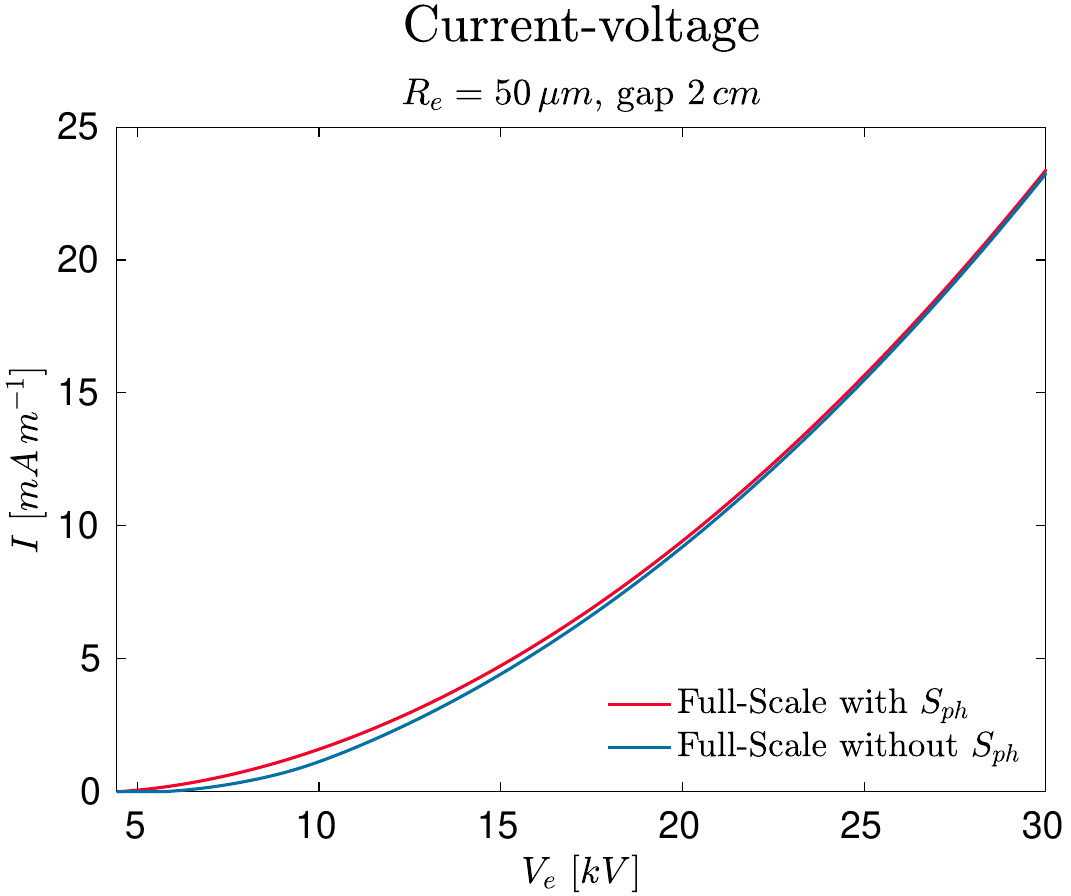} %
    \caption{Current-voltage curve obtained from full-scale simulations. Emitter radius $50\,\mu\text{m}$, gap $2\, \text{cm}$.}
    \label{fig:iv_curve_high_current}
\end{figure}

We have noticed that, when the emitter radius is decreased and the applied voltage increases, the role of photo-ionization becomes less and less important in predicting the current.
Figure~\ref{fig:iv_curve_high_current} shows the current-voltage curve for an emitter of radius $R_e = 50\, \mu\text{m}$ and a gap $L=2\, \text{cm}$ with and without photo-ionization. When the applied voltage is above $15\,\text{kV}$ the difference in the computed current is negligible. However, there is still a difference in the neighborhood of the discharge onset, that is around $5\,\text{kV}$. We remark also, as before, that the onset is delayed without photo-ionization.\\

\subsection{Ionization region thickness}
\label{sec: IonizationLength}

\begin{figure}[H]
    \centering
        \includegraphics[width=0.6\textwidth]{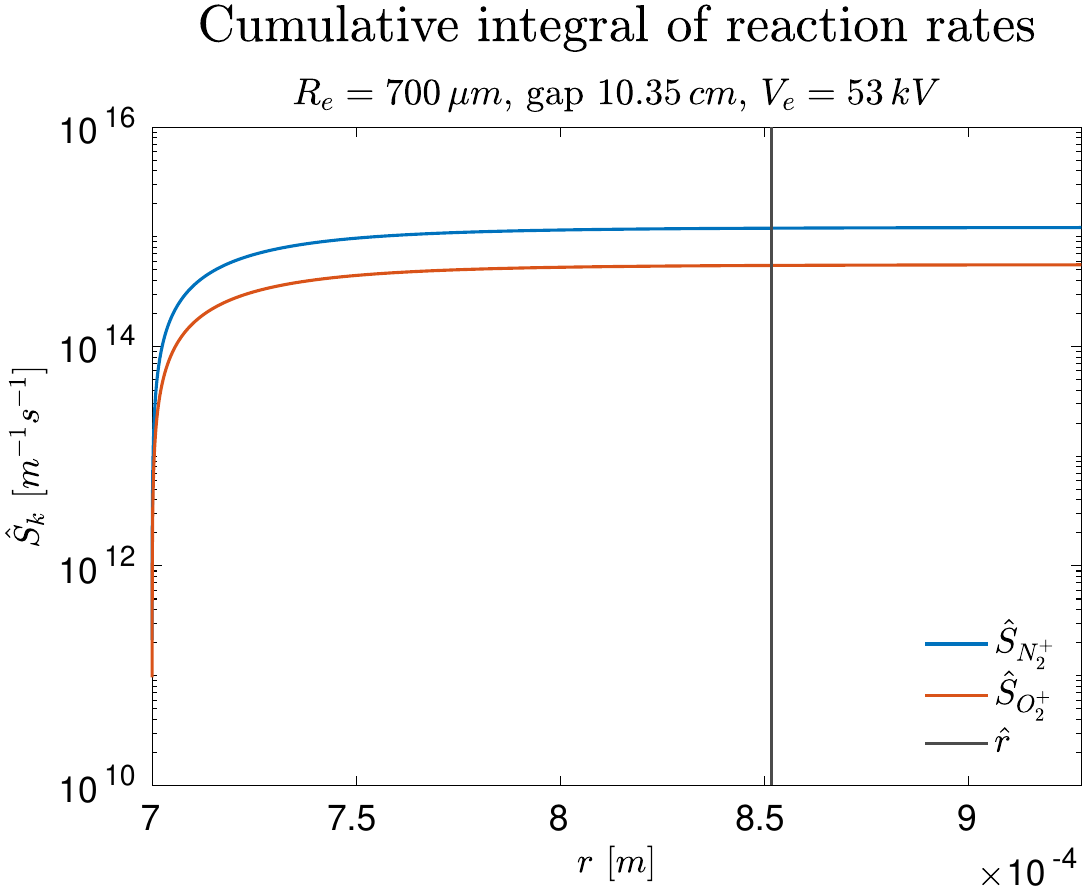} %
        \caption{Cumulative integral of the reaction rates, applied voltage $V_e=53\,\text{kV}$.}
        \label{fig: cumulativeIntegral}
    \end{figure}

\begin{figure}[H]
        \centering
        \includegraphics[width=0.7\textwidth]{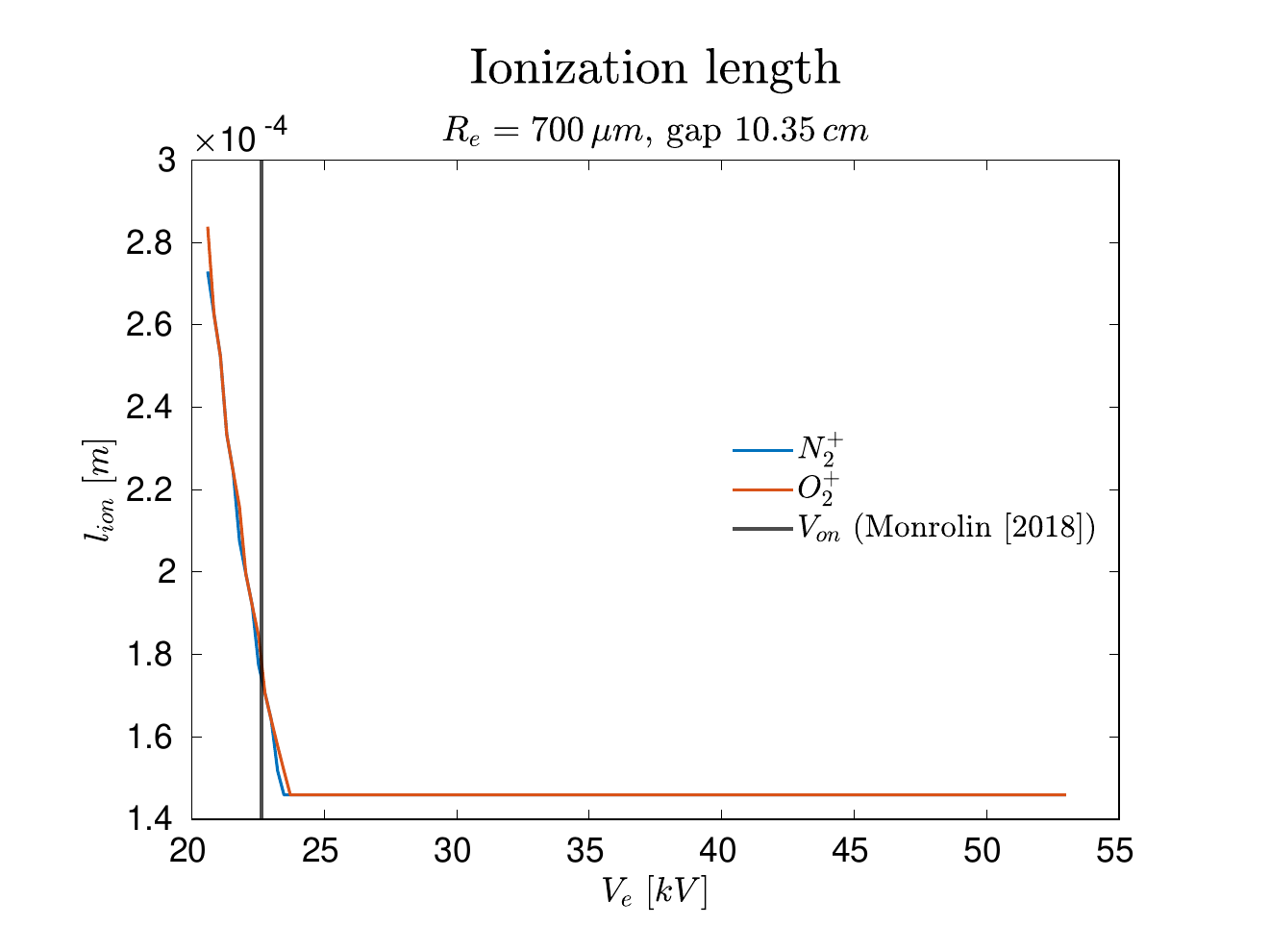} %
        \caption{Ionization length as a function of the applied potential for positive ions.}
        \label{fig: ionL_vs_V}
\end{figure}

As already pointed out, in corona discharges the strong gradient of the electric field allows the distinction between an ionization and a drift region. To the authors' best knowledge, there is not a unique criterion to identify the length of the plasma region. Therefore we propose a strategy to identify where it is possible to reasonably locate the end of the ionization layer and thus to compute the plasma region length. Our idea is based on the phenomenological behavior of corona discharges, assuming a clear distinction between the ionization and the drift zone.
By definition, the ionization layer is where the greatest part of ionizing events by chemical reactions takes place.
We define a cumulative integral of the $k^{th}$ species source term as:
\begin{align}
    \hat{S}_k(r)=\int_{R_e}^{r} S_k r' dr' \qquad k=N_2^+,O_2^+
    \label{eq:cumulativeint}
\end{align}
The integral $\hat{S}_k(r)$ is computed for every relevant charged species that will survive in the drift region, namely the positive ions, that for the Parent model are $N_2^+$ and $O_2^+$.
Remark that, because of symmetry, it is unnecessary to integrate over the remaining coordinates of the cylindrical reference system.
The cumulative integral is equal to the total generation or destruction of species $k$ in the domain, per unit time and area, when $r=R_c$.
We can find the coordinate of the end of the ionization layer by considering the point at which the cumulative integral reaches a certain threshold. A low threshold will underestimate the thickness of the ionization region, a too high one will overestimate it. An example of cumulative integral is presented in Fig.~\ref{fig: cumulativeIntegral} for an emitter radius $R_e=700\,\mu\text{m}$, a gap $L=10.35\,\text{cm}$ and an applied voltage of $53000\, \text{V}$. This case corresponds to the highest applied voltage of Fig.~\ref{fig:iv_curve_validation} in Sec.~\ref{sec:validation}. The cumulative integrals of $N_2^+$ and $O_2^+$ both increase very quickly near the emitter, reach a maximum and slightly decrease afterwards in the drift region, where there is a practically negligible neutralization of charged species.
The value $\hat{r}$ at which $\hat{S}_k$ reaches $98\%$ of its first maximum is considered as the end of the ionization layer. With this procedure we obtain two values of $\hat{r}$, one for $N_2^+$ and one for $O_2^+$. The end of the ionization region, and therefore the ionization length $l_{ion}=\hat{r}-R_e$, is then identified with the maximum between these two values. Once the corona discharge is ignited, the two values of the computed ionization lengths are close one to the other, as shown in Fig.~\ref{fig: ionL_vs_V}, thus supporting the validity of our definition of the ionization region length. In principle it is necessary to compute the ionization length for every applied voltage greater than the onset one. However, as we notice from Fig.~\ref{fig: ionL_vs_V}, the ionization length quickly tends to an asymptotic value. Therefore, in the following, given a range of voltages for a certain geometry, we retain the value of ionization length that corresponds to the last simulated applied voltage.


\subsection{Injection law}
\label{sec: LookUpTable}

\begin{figure}
    \centering
     \includegraphics[width=0.55\textwidth]{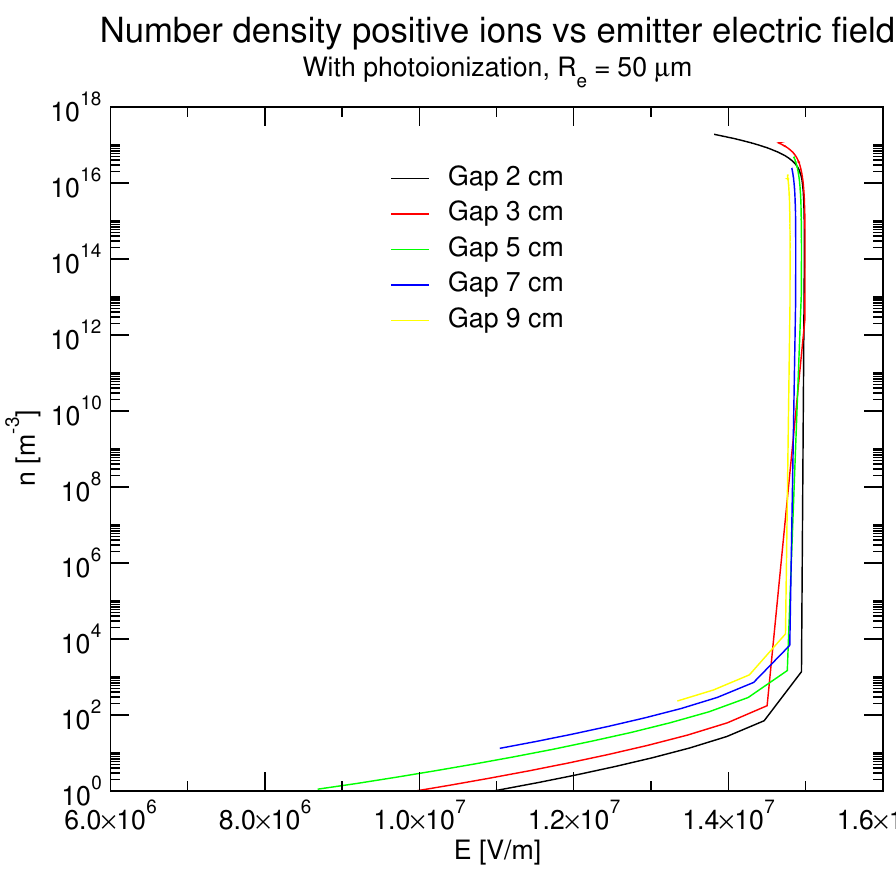}
    \caption{Number density of positive ions versus electric field at the emitter surface, with photo-ionization.}
    \label{fig:np_vs_elec}   
\end{figure}

\begin{figure}
\centering
        \includegraphics[width=0.55\textwidth]{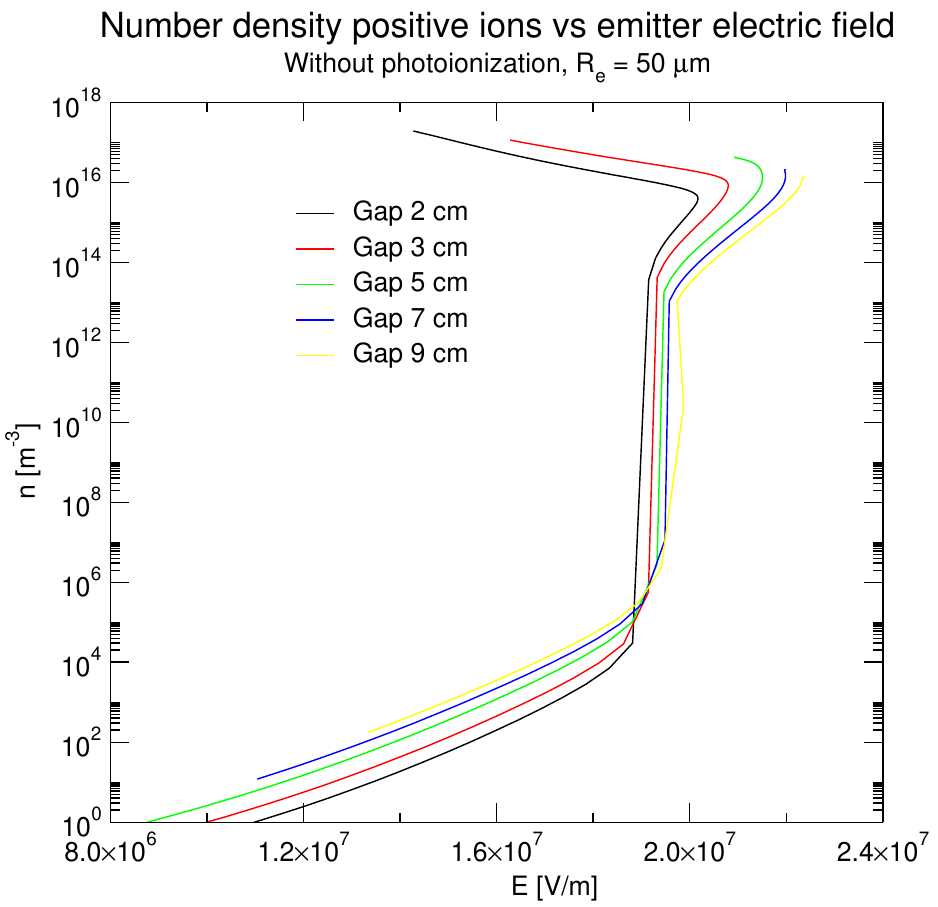}
        \caption{Number density of positive ions versus electric field at the emitter surface, without photo-ionization.}
    \label{fig:np_vs_elec_noph}
\end{figure}

The main goal of this work is to provide a reliable way of exploiting one dimensional results to build boundary conditions for more complex configurations.
To accomplish this task we need to sample the positive charge density at the end of the ionization region as function of the electric field.
In the macro-scale, or drift region, computations, the ionization region is assumed to be vanishingly small, i.e.\ the macro-scale boundary conditions are applied at the surface of the emitter electrode, instead of the edge of the ionization layer. For this reason, we use the electric field at the emitter surface as a parameter to fit the number density of positive ions. This approach is very similar to classical boundary layer theory~\cite{white}, where the outer edge of the boundary layer (the ionization region for us) corresponds to the wall surface of the external flow (the drift region for us).
We have noticed that photo-ionization strongly influences the relation between charge density and electric field at the edge of ionization region (see Figs.~\ref{fig:np_vs_elec} and~\ref{fig:np_vs_elec_noph}), even for cases, like the one shown in Fig.~\ref{fig:iv_curve_high_current} of Sec.~\ref{sec:validation},  where the current is only marginally affected by photo-ionization. Therefore we performed full-scale computations with and without photo-ionization and differentiate the fitting procedure between the two cases.\\
We have carried out several simulations varying the emitter radius and the gap length. We considered five different emitters with radius $R_e = 700, 500, 300, 100, 50\, \mu\text{m}$ and five gaps $L = 9, 7, 5, 3, 2\, \text{cm}$. The computations are performed with the full-scale model, presented in Sec.~\ref{sec:fullscalemod} and validated in Sec.~\ref{sec:validation}. The parameters for the boundary condition of the macro-scale model (positive charge number density and electric field) are found from these full-scale results.\\
For the computations with photo-ionization, Fig.~\ref{fig:np_vs_elec} clearly shows that, once the discharge is ignited, the relation between the number density of positive ions at the edge of the ionization region and electric field at the emitter surface is quite insensitive to the gap length. 
A Kaptzov-like behavior is observed: once the discharge is ignited the value of the electric field remains nearly constant, with a huge increase in the density of positive ions. 
The results of Fig.~\ref{fig:np_vs_elec} are computed for an emitter with $R_e=50\,\mu m$, but the same qualitative behavior is found for the other combinations of emitter radius and gap. Therefore we consider only the biggest gap, namely $L=9\, \text{cm}$, for every emitter radius.\\
For computations without photo-ionization, the differences among the various gap lengths are more important, as shown in Fig.~\ref{fig:np_vs_elec_noph} and
the Kaptzov like behavior is much less evident.
In particular for gaps of $2\, \text{cm}$ and $3\, \text{cm}$ the link between electric field and number density is not a one-to-one function anymore. This phenomenon happens at high applied voltages, when the discharge is probably very close to spark transition. We are aware that further investigations are needed before we can conclude if this behavior can be real or it is only an artifact of our modeling assumptions. As a consequence we have discarded the results for the small gaps and have taken into account only the biggest gap also for the fitting procedure without photo-ionization.

\begin{figure}[H]
    \centering
        \includegraphics[width=0.6\textwidth]{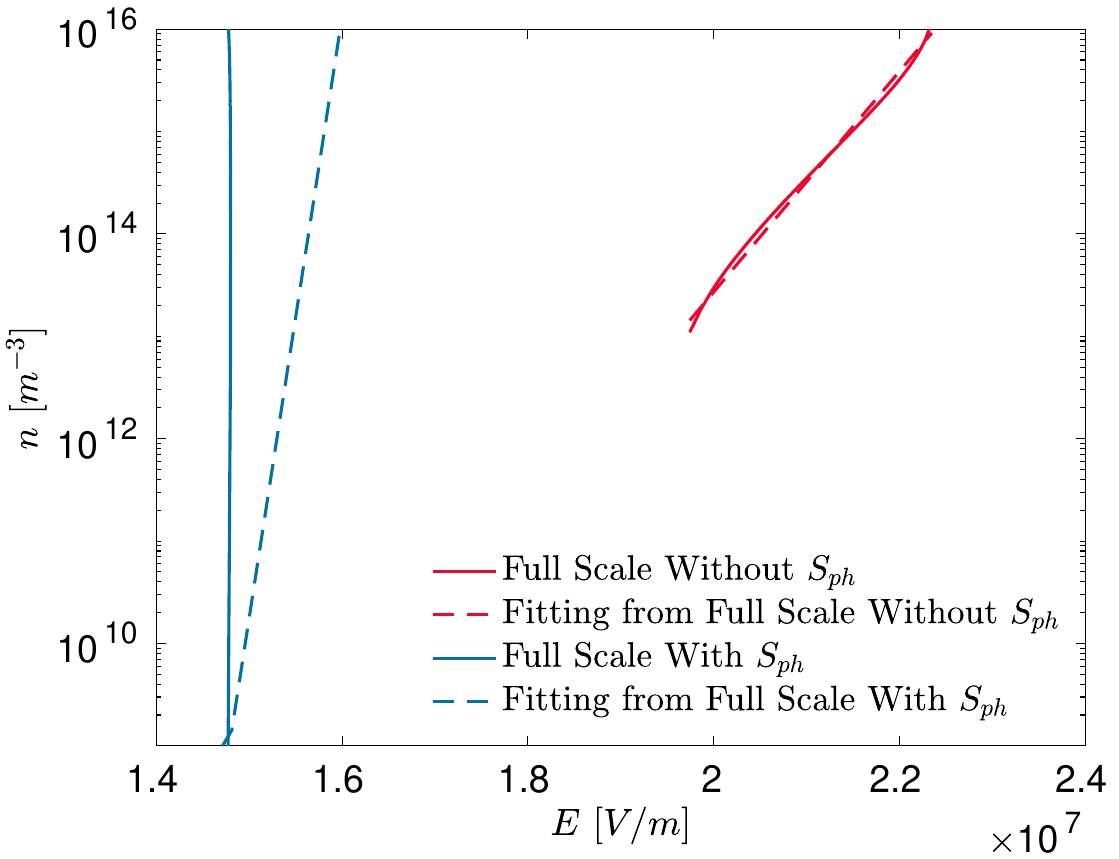} %
        \caption{Fitting: $R_e=50\,\mu\text{m}$, gap $L=9\,\text{cm}$. \label{fig:fitting50}}
    \end{figure}

\begin{figure}
        \centering
        \includegraphics[width=0.6\textwidth]{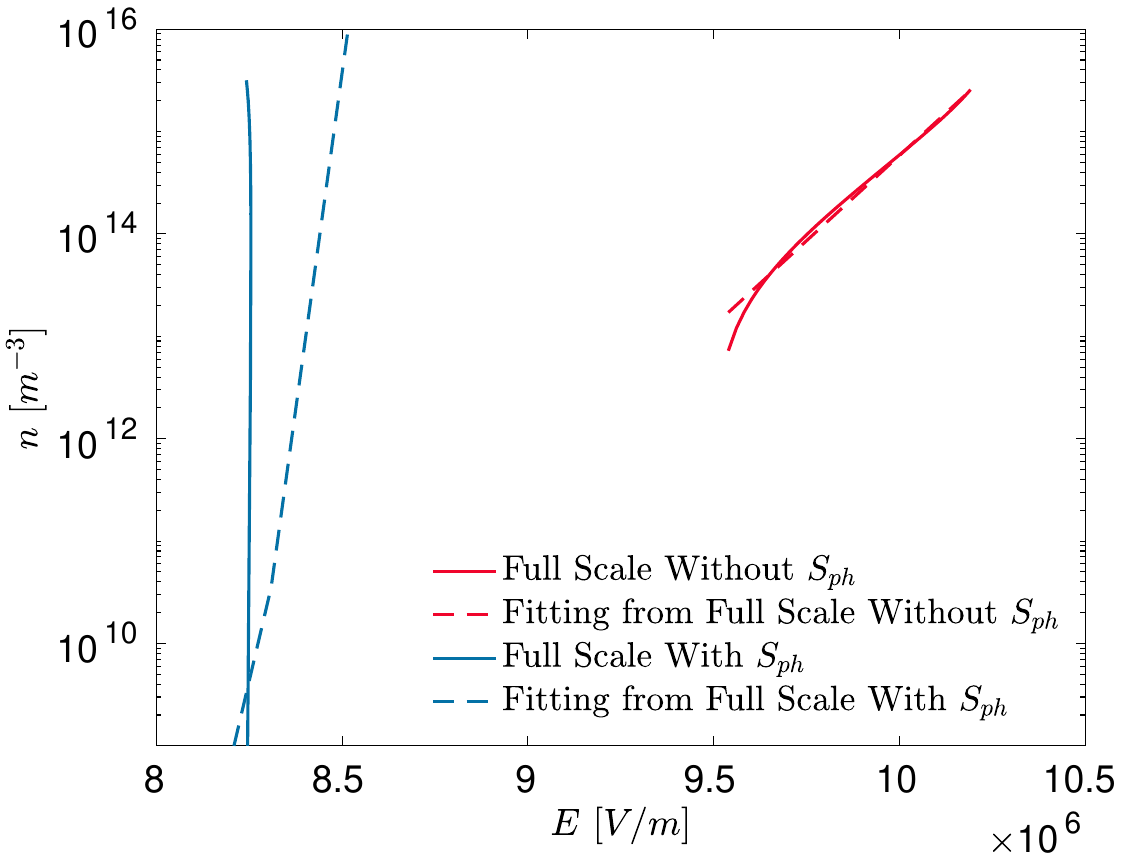} %
        \caption{Fitting: $R_e=300\,\mu\text{m}$, gap $L=9\,\text{cm}$. \label{fig:fitting300}}
    \end{figure}

\begin{figure}
        \centering
        \includegraphics[width=0.6\textwidth]{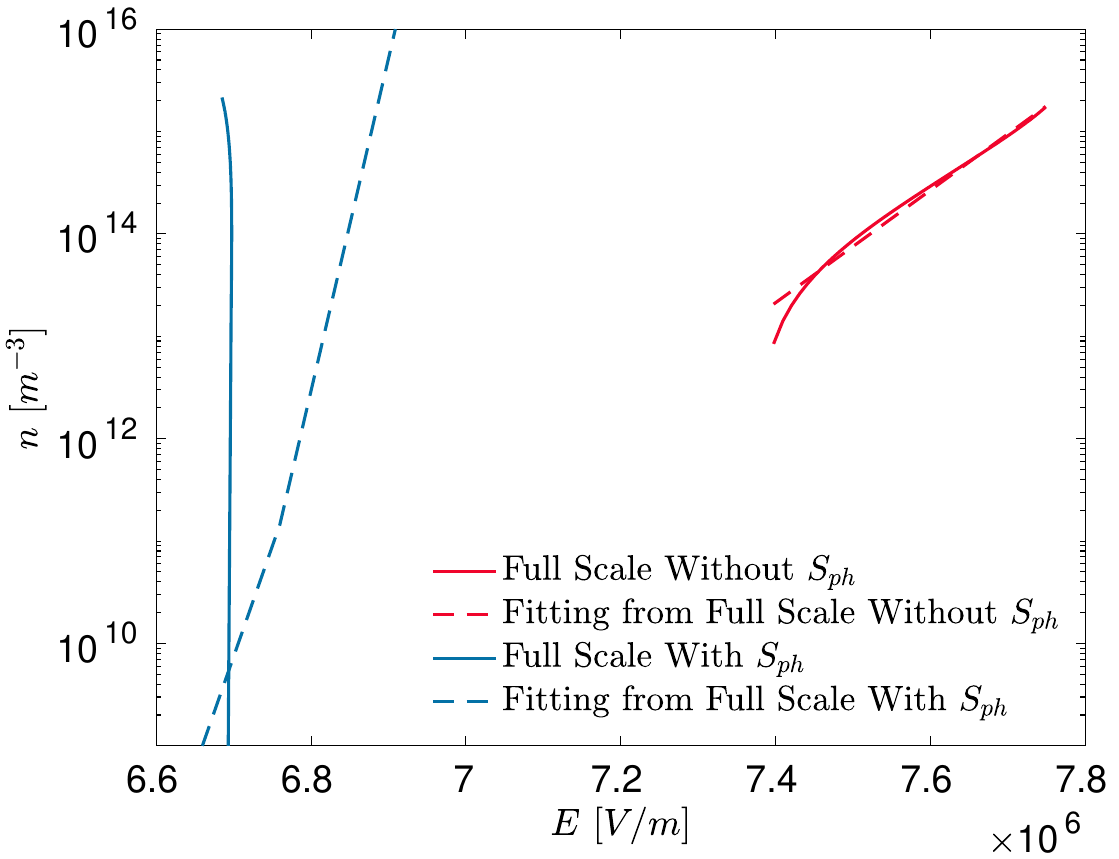} %
        \caption{Fitting: $R_e=700\,\mu\text{m}$, gap $L=9\,\text{cm}$. \label{fig:fitting700}}
   
\end{figure}

We fit the relation between the positive ions density and the electric field with an exponential law of the form:
\begin{equation}
    n_p(E)= n_{ref}\exp\left(\frac{E-\Bar{E}_{on}}{E_{ref}}\right) 
    \label{eq:fitting_exp}
\end{equation}
The rationale is that this form has already been proven to work well and to ensure the numerical stability of the non linear solver of the macroscopic model~\cite{Cagnoni}.\\
Below the onset, when the discharge is not yet started, the density of positive ions is very small and the current is negligible. It is pointless to fit the relation between density and electric field for this condition, therefore the density is simply kept equal to a suitable constant value. In the self-sustained regime the density profile approximated by Eq.~\ref{eq:fitting_exp} can be seen as a linear regression in the log-log plane.
Taking the logarithm on both sides of Eq.~\ref{eq:fitting_exp} it follows:
\begin{equation}
   \ln(n_p/n_{ref})= \frac{E-\Bar{E}_{on}}{E_{ref}} = mE + q 
\end{equation}
where $m=1/E_{ref}$ and $q=-\Bar{E}_{on}/E_{ref}$.\\
$\Bar{E}_{on}$ can be interpreted as a kind of onset electric field, although it should not be confused with the physical quantity $E_{on}$ discussed in Sec.~\ref{sec:validation}. It can coincide numerically with the true $E_{on}$, but it is better to see it as a fitting parameter. $n_{ref}$ is the value of number density when $E=\Bar{E}_{on}$. $E_{ref}$ will determine the slope of the linear regression. The interpretation of this parameter is different according to the full scale model adopted.\\
The continuous curves of Figs.~\ref{fig:fitting50}, \ref{fig:fitting300} and \ref{fig:fitting700} show the number density of positive ions as a function of the electric field, once the discharge is ignited. With photo-ionization the curves follow quite closely the Kaptzov hypothesis, hence the electric field remains almost constant. This behavior causes troubles in the numerical solution of the macroscopic model, because a small change in the electric field would correspond to a huge variation in ions number density: here $E_{ref}$ controls the slope of the number density and the stability of the numerical solution. A small $E_{ref}$ yields a steeper slope but at the same time reduces the beneficial effect of the exponential law in terms of stability. Therefore $E_{ref}$ needs to be a compromise between accuracy of fitting and numerical stability. We empirically found that $E_{ref}$ should be set between $10^{-3}\, \Bar{E}_{on}$ and $10^{-2}\, \Bar{E}_{on}$; further details are given in Appendix~\ref{sec:appa}.\\
In Sec.~\ref{sec:validation}, Fig.~\ref{fig:iv_curve_high_current}, we have shown that, for small radius emitters, the photo-ionization plays a marginal role in the current-voltage characteristic of the discharge, for a sufficiently high applied voltage. However the model produces profiles of the number density versus the electric field at the emitter that are different from the ones predicted with photo-ionization, as can be noticed by Figs.~\ref{fig:np_vs_elec} and~\ref{fig:np_vs_elec_noph}.
Indeed, without photo-ionization, the number density shows initially a Kaptzov-like behavior after the onset, but when the number density is above $10^{13}$, the behavior deviates from the Kaptzov one and, for gaps greater than $5\, cm$, the increase of positive charges with electric field is smoother. In this case $E_{ref}$ is a parameter that permits to find the correct slope in the $n_p-E$ plane.\\
In Figs.~\ref{fig:fitting50}, \ref{fig:fitting300} and \ref{fig:fitting700} the exponential fitting of the boundary condition is reported considering various emitter radii, with and without photo-ionization. Without photo-ionization a systematically higher electric field and therefore applied voltage is needed to reach the same value of ions number density produced with photo-ionization. The consequence in the macroscopic model is that, for the same applied voltage, the current will be under-estimated because the onset is delayed, as shown in Figs.~\ref{fig:iv_curve_validation} and~\ref{fig:iv_curve_high_current} for the full-scale model.
For low applied voltages, when the discharge has not yet started, the error of the described fitting will be considerable. However, because we are interested only in the self-sustained regime of the corona, this error is not of practical relevance.
The values of the parameters $n_{ref}$, $\Bar{E}_{on}$ and $E_{ref}$ used to obtain the fitting curves of Figs.~\ref{fig:fitting50}, \ref{fig:fitting300} and \ref{fig:fitting700} are reported in Table~\ref{tab: fitting_parameters}. We observe that, with photo-ionization, $\Bar{E}_{on}$ is very close to the values of the physical $E_{on}$ shown in Table~\ref{tab: OnsetErrors}; without photo-ionization the difference between the two quantities is quite larger.
\begin{table}[H]
    \footnotesize
    \centering
     \caption{Values of the parameters $n_{ref}$, $\Bar{E}_{on}$ and $E_{ref}$ used to obtain the fitting curves of Figs.~\ref{fig:fitting50}, \ref{fig:fitting300} and \ref{fig:fitting700}}
    \begin{tabular}{ccccc}
       Emitter Radius [$\mu\text{m}$] & 
       $S_{ph}$ & 
       $\Bar{E}_{on}$ [$\text{V/m}$]& 
       $E_{ref}$ [$\text{V/m}$]& 
       $n_{ref}$ [$\text{m}^{-3}$] \\ 
        \hline
        \rule{0pt}{1.5em}
        $50$    &Yes &  $1.4788\;10^7$   &$7.3938\;10^4$     & $10^9$\\[0.5em]
        $50$    &No  &  $1.5886\;10^7$   &$4.0299\;10^5$     & $10^9$\\[0.5em]
        $300$   &Yes &  $8.2503\;10^6$   &$1.6501\;10^4$     & $10^9$\\[0.5em]
        $300$   &No  &  $8.2679\;10^6$   &$1.3044\;10^5$     & $10^9$\\[0.5em]
        $700$   &Yes &  $6.693\;10^6$   &$ 1.3386\;10^4$    & $10^9$\\[0.5em]
        $700$   &No  &  $ 6.6121\;10^6$  &$ 7.9044\;10^4$    & $10^9$\\[0.5em]
    \end{tabular}
    \label{tab: fitting_parameters}
 
\end{table}


The blue curves in Figs.~\ref{fig:fitting50}, \ref{fig:fitting300} and \ref{fig:fitting700} compare the injection law computed with the full-scale model and the one predicted by Eq.~\ref{eq:fitting_exp} with the coefficients of Table~\ref{tab: fitting_parameters}, when photo-ionization is included in the full-scale model. The difference in the slope visually represents the relaxation of Kaptzov's hypothesis due to the fitting formula.
Despite the apparently non-optimal fitting, we show in Secs.~\ref{sec: Macroscopic} and \ref{sec: Macroscopic2D} that the macroscopic model provides good predictions of current and ions number density; and in Appendix~\ref{sec:appa} that the results are quite insensitive to the slope of the fitting curve. The reason is that the electric field cannot increase arbitrarily: an increase of the electric field would lead to an exponential growth of ion concentration, which in turn would stabilize the value of the electric field through a feedback effect.
With Kaptzov hypothesis, the field is precisely clamped to a specific value $E_{on}$; here, instead, it is constrained by the exponential fitting of Eq.~\ref{eq:fitting_exp} to remain within a neighborhood of $E_{on}$.


\section{Macroscopic Simulations}
\label{sec: Macroscopic}

\begin{figure}
    \centering
        \includegraphics[width=0.65\textwidth]{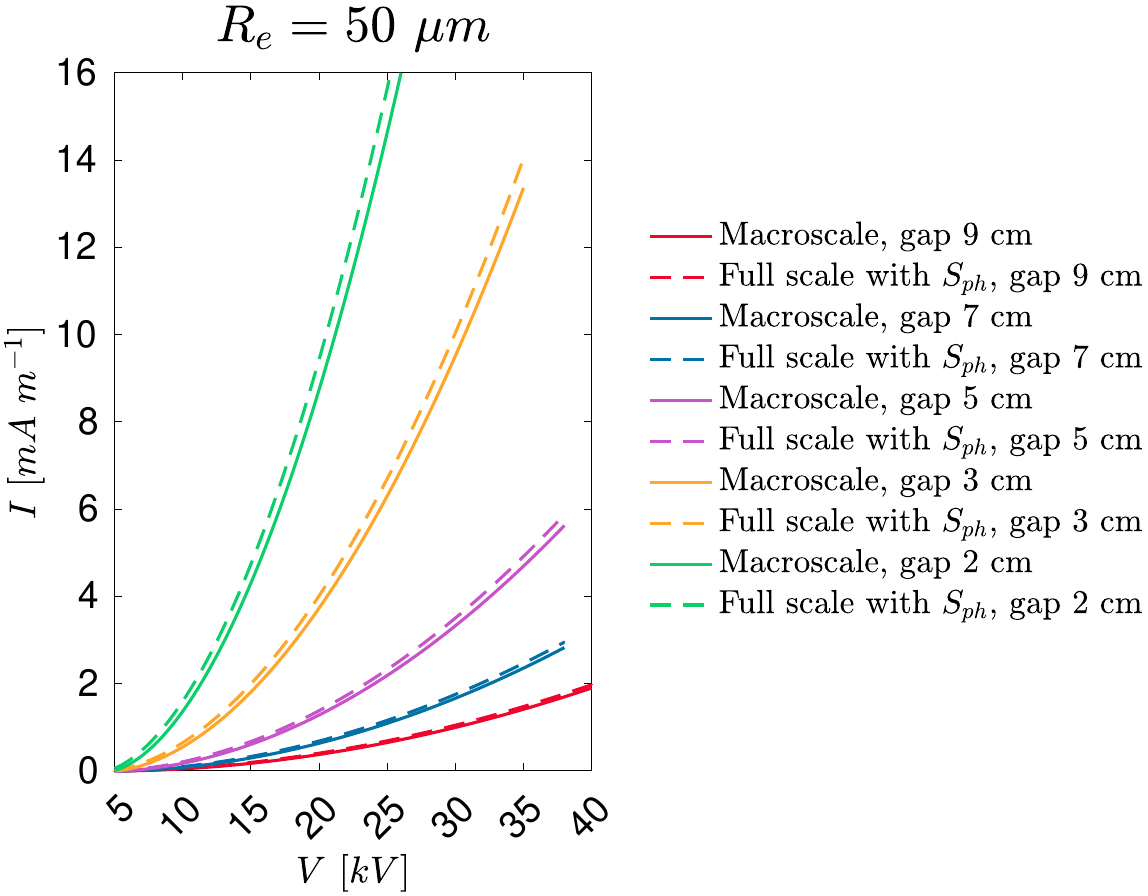}
        \caption{Current-voltage curves for $R_e=50\,\mu\text{m}$. Macroscopic simulations versus full-scale ones.
        \label{fig:iv_curve_macrofull50}}
    \end{figure}

    \begin{figure}
        \centering
        \includegraphics[width=0.65\textwidth]{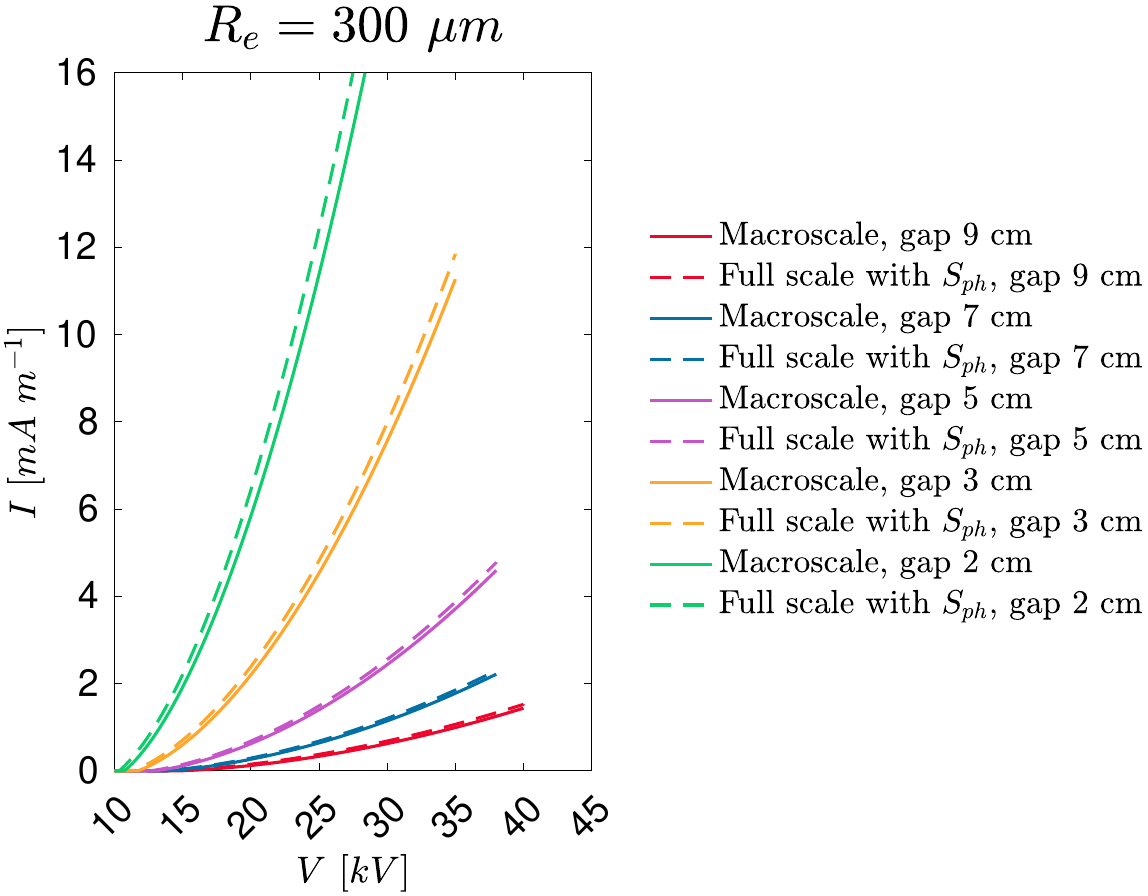}
        \caption{Current-voltage curves for $R_e=300\,\mu\text{m}$. Macroscopic simulations versus full-scale ones.
        \label{fig:iv_curve_macrofull300}}
    \end{figure}

    \begin{figure}
        \centering
        \includegraphics[width=0.65
        \textwidth]{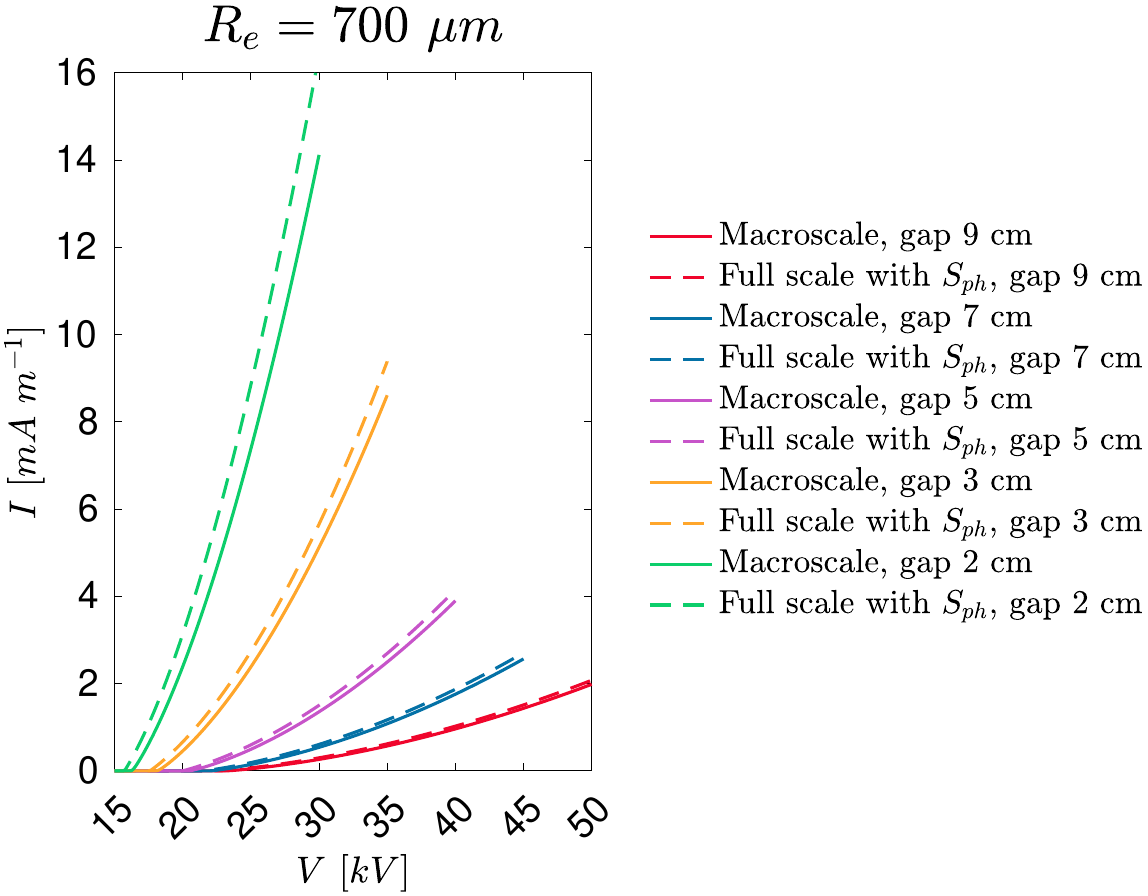}
         \caption{Current-voltage curves for $R_e=700\,\mu\text{m}$. Macroscopic simulations versus full-scale ones.
         \label{fig:iv_curve_macrofull700}}
    \end{figure}


In this Section we compare the results of the macro-scale model, supplemented with the injection law explained in Sec.~\ref{sec: LookUpTable}, with the full-scale computations. We emphasize that, for each emitter, only one full-scale simulation is required to fit the boundary condition for a variety of gap lengths in the macro-scale model. \\
Figures~\ref{fig:iv_curve_macrofull50}, \ref{fig:iv_curve_macrofull300} and \ref{fig:iv_curve_macrofull700} show the current-voltage curves computed by the macro-scale model for three different emitter radii ($R_e=50\,\mu\text{m}$, $R_e=300\,\mu\text{m}$, $R_e=700\,\mu\text{m}$) against those computed with the full-scale one, with photo-ionization. 
As a general trend we observe that the agreement is excellent for the biggest gaps, which is not surprising because the fitting procedure of Sec.~\ref{sec: LookUpTable} is carried out for a gap
of $9\, \text{cm}$, and is acceptable for the smallest gaps of $2\, \text{cm}$ and $3\, \text{cm}$.
The maximum error is found for the combination of smallest gap and biggest emitter and it is about $15\,\%$. There are a couple of reasons that justify this behavior. First of all we remark that the current, with the same applied voltage, decreases with respect to gap length. The reason is simple, the number density of positive ions at the emitter decreases too for increasing gap length. An inspection of Figs.~\ref{fig:fitting50}, \ref{fig:fitting300} and \ref{fig:fitting700} shows clearly that the discrepancy between the true $n_p-E$ curve and its fitting increases with $n_p$. Therefore small gap lengths, with higher $n_p$, amplify the discrepancy between full-scale and macro-scale computations, even if the maximum error is still acceptable. The agreement could be improved increasing the slope of the fitting curve, i.e.\ with a smaller $E_{ref}$, but at the expense of numerical stability, as explained in Sec.~\ref{sec: LookUpTable}. 
Another reason is that some deviations from the Kaptzov-like behavior can arise for small gaps and high applied voltage: indeed this is shown in Fig.~\ref{fig:np_vs_elec} for $2\,\text{cm}$ and $3\, \text{cm}$ gaps. In the context of negative corona, small deviations from the Kaptzov hypothesis have been found in high current regime~\cite{NegativeCorona}. A detailed study of these phenomena is beyond the scope of the present paper and it is postponed to future studies.

\begin{figure}
        \centering
        \includegraphics[width=0.6\textwidth]{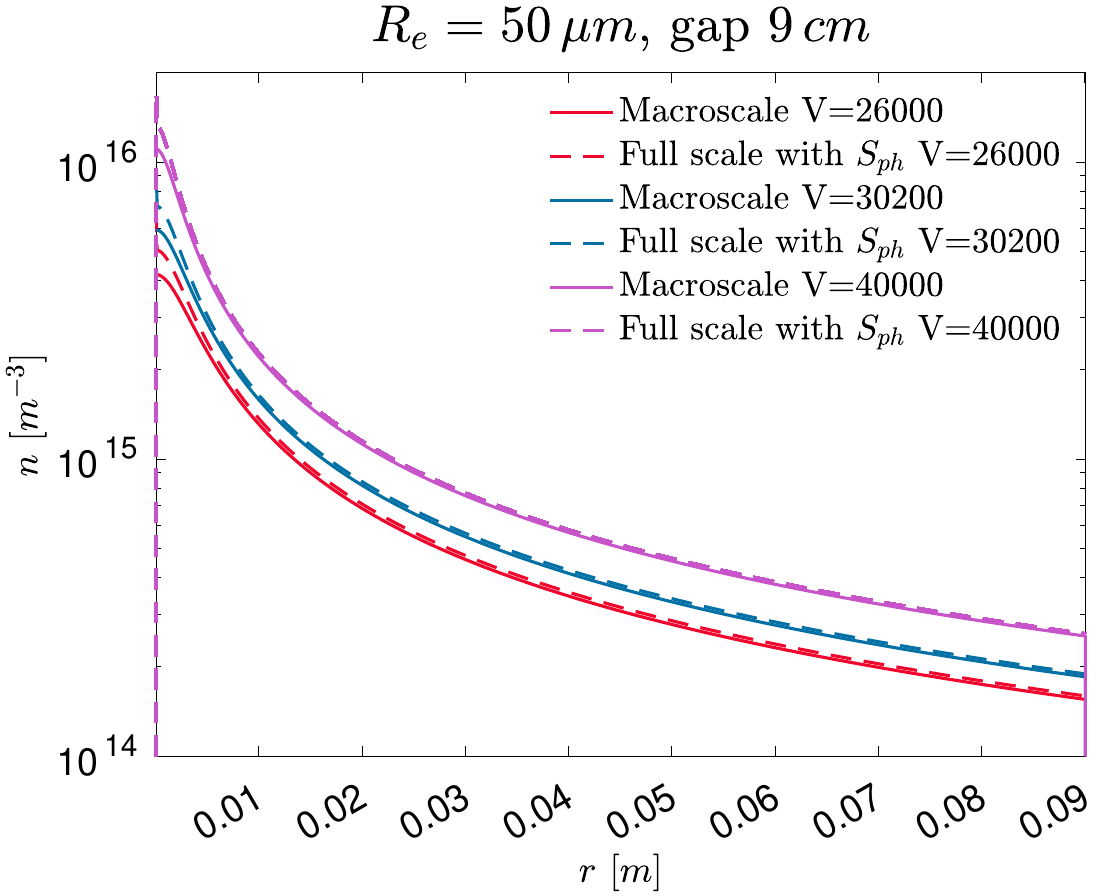} %
    \caption{Number density of positive ions in function of applied voltage. Macro-scale model versus full-scale one. $R_e=50\,\mu\text{m}$, gap length $L= 9\, \text{cm}$.
    \label{fig:density9_50}}
\end{figure}

\begin{figure}
\centering
        \centering
        \includegraphics[width=0.6\textwidth]{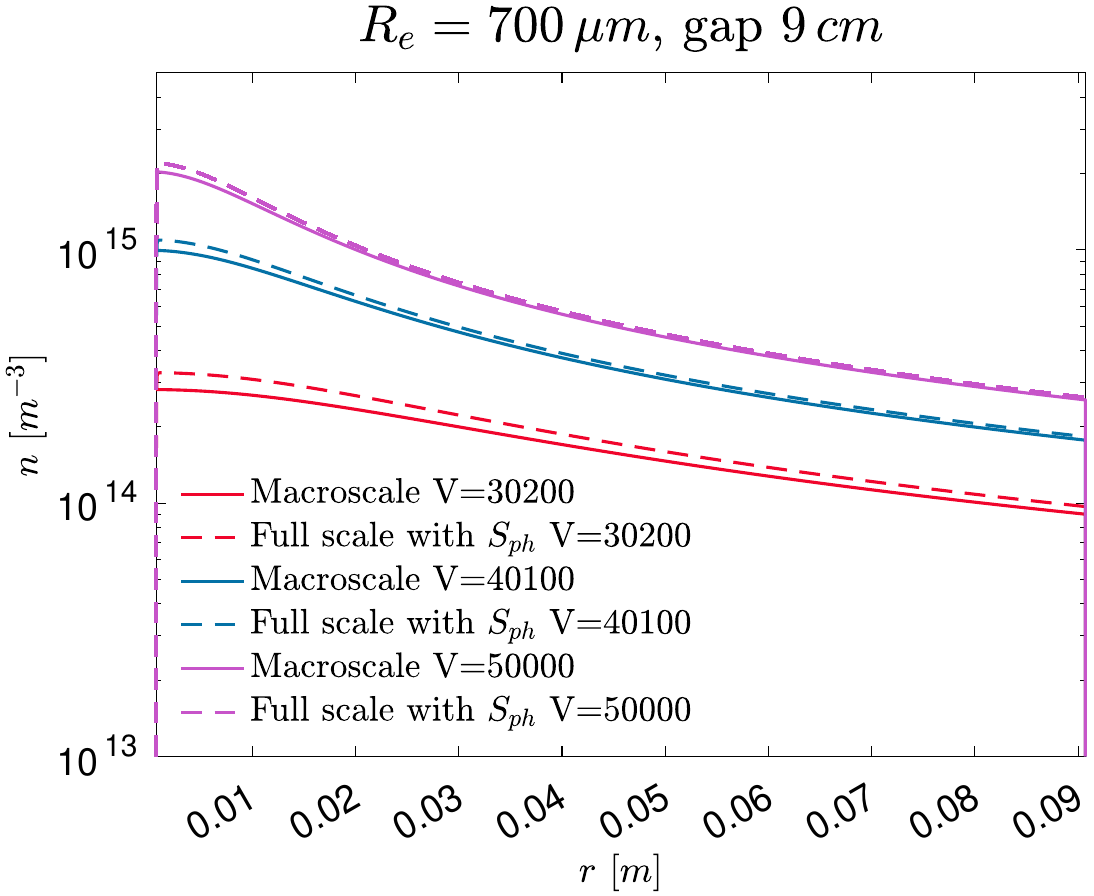} %
        \caption{Number density of positive ions in function of applied voltage. Macro-scale model versus full-scale one. $R_e=700\,\mu\text{m}$, gap length $L= 9\, \text{cm}$.
        \label{fig:density9_700}}
\end{figure}

\begin{figure}
        \centering
        \includegraphics[width=0.6\textwidth]{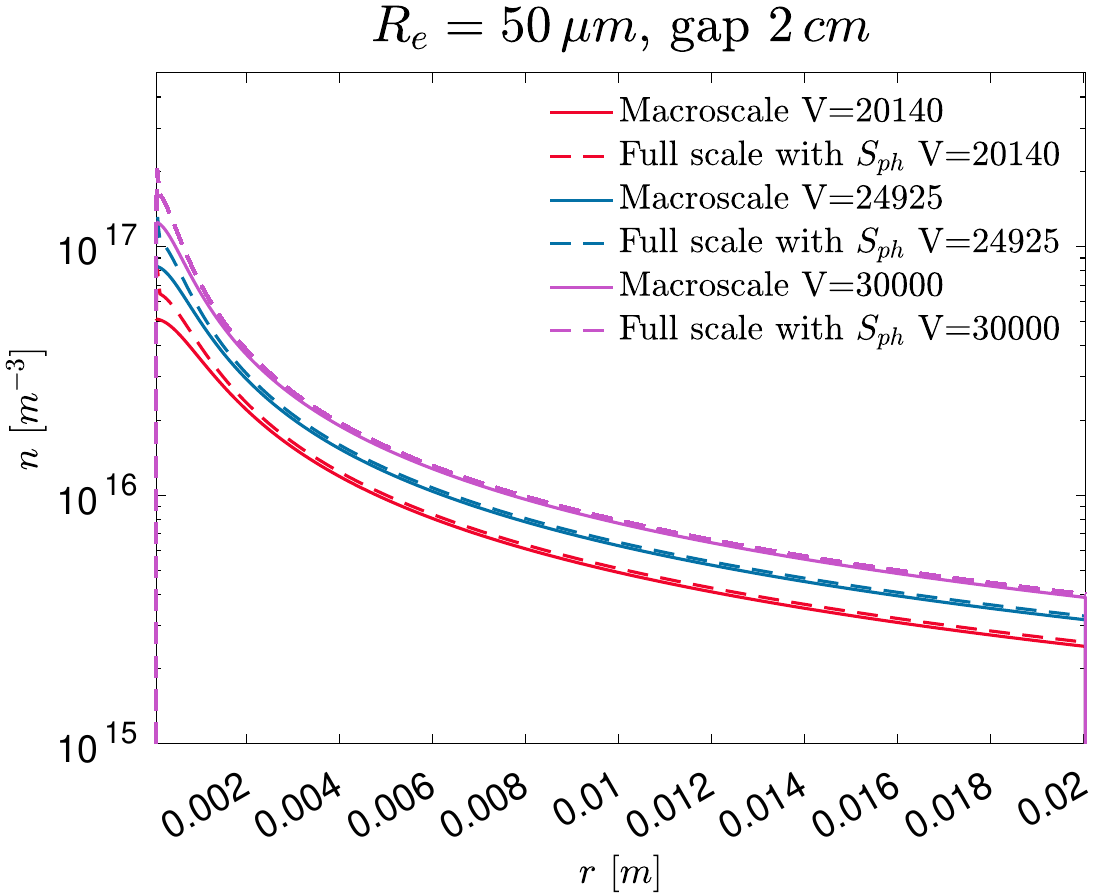} %
        \caption{Number density of positive ions in function of applied voltage. Macro-scale model versus full-scale one. $R_e=50\,\mu\text{m}$, gap length $L=2\, \text{cm}$.
        \label{fig:density2_50}}

\end{figure}

\begin{figure}
    \centering
        \centering
        \includegraphics[width=0.6\textwidth]{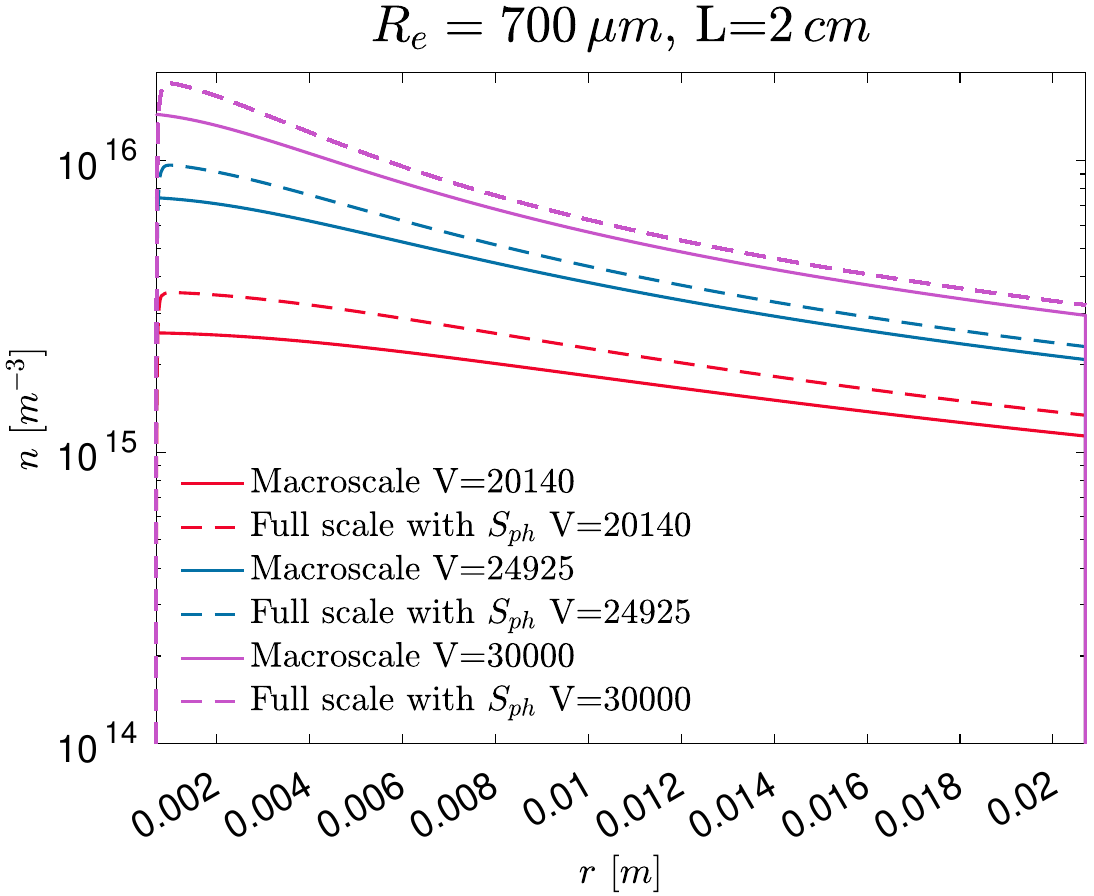} %
        \caption{Number density of positive ions in function of applied voltage. Macro-scale model versus full-scale one. $R_e=700\,\mu\text{m}$, gap length $L=2\, \text{cm}$.
        \label{fig:density2_700}}
\end{figure}

An inspection of Figs.~\ref{fig:density9_50} and~\ref{fig:density9_700} shows that for a gap $L=9\,\text{cm}$ the density profile obtained with the macroscopic model is in very good agreement with the full-scale one, since the boundary condition is derived from this geometry. 
Differences between the two models are amplified for a gap $L=2\,\text{cm}$; the macroscopic model tends to underestimate the charge injected in the domain, as shown in Figs.~\ref{fig:density2_50} and~\ref{fig:density2_700}. Again this is expected since we are limiting the slope of the fitting curve between the number density of positive ions and the electric field at the emitter surface. We remark again that the less accurate configuration is for the smallest gap and the biggest emitter radius, as shown by Fig.~\ref{fig:density2_700}.
In the field of ionic propulsion emitter radii are usually below $100\,\mu\text{m}$~\cite{Belan} and therefore, as we remark from Figs.~\ref{fig:density9_50} and~\ref{fig:density2_50}, macro-scale results can be expected to be quite close to full-scale ones.

\begin{figure}
    \centering
        \includegraphics[width=0.6\textwidth]{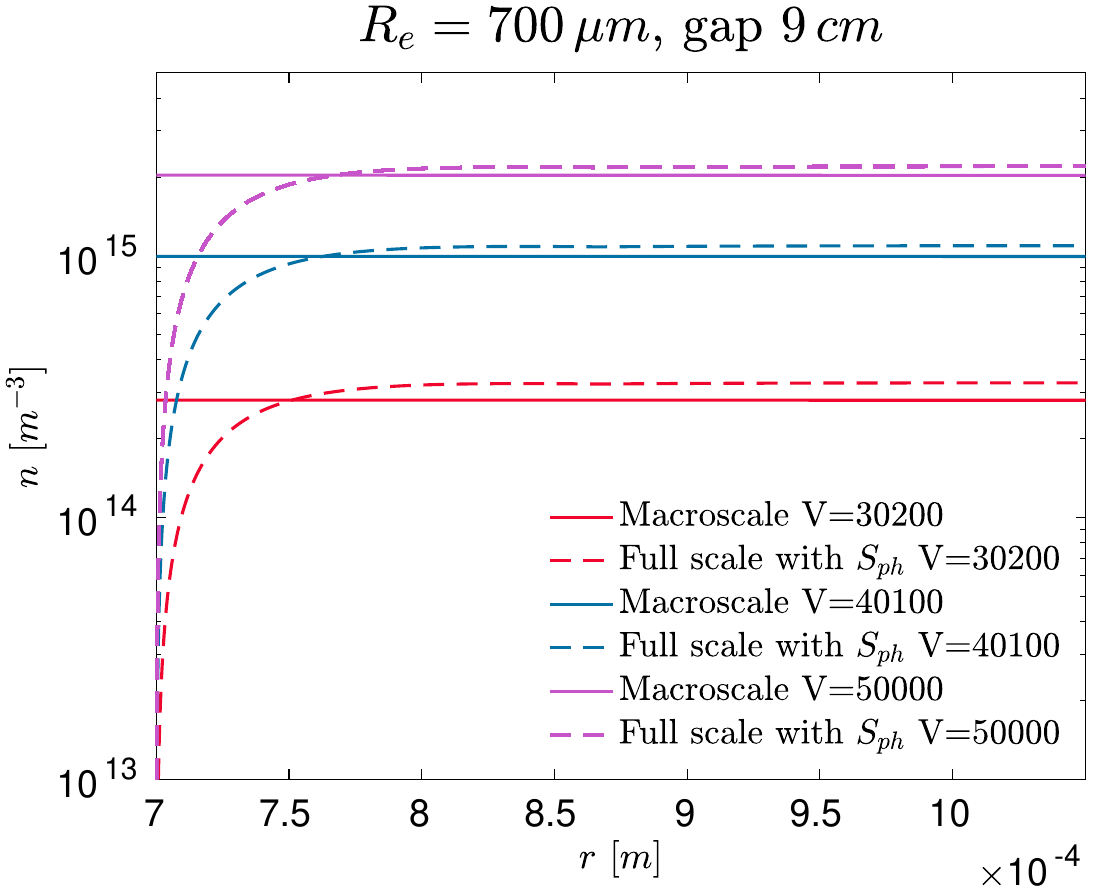} %
        \caption{Number density of positive ions in function of applied voltage. Zoom of the emitter region. $R_e=700\,\mu\text{m}$, gap length $L=9\, \text{cm}$.
        \label{fig:zoom_density9_700}}
\end{figure}

\begin{figure}
        \centering
        \includegraphics[width=0.6\textwidth]{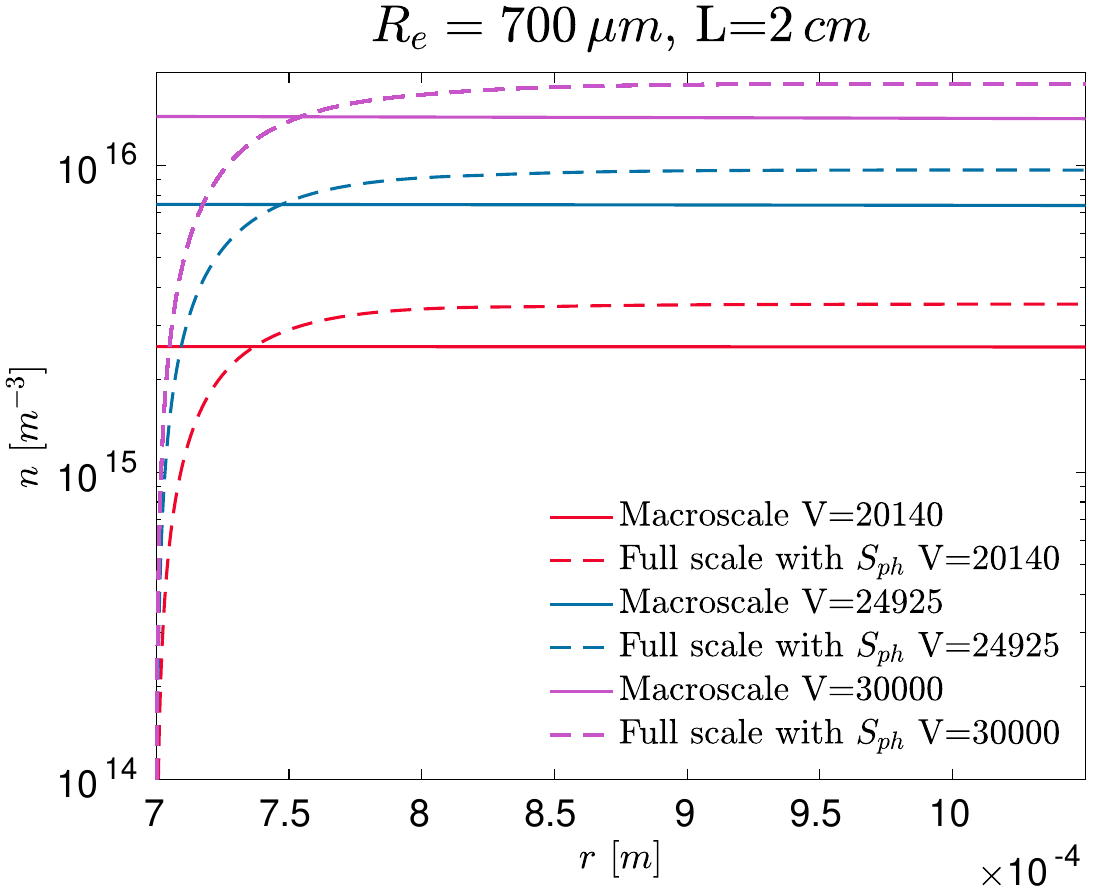} %
        \caption{Number density of positive ions in function of applied voltage. Zoom of the emitter region. $R_e=700\,\mu\text{m}$, gap length $L=2\, \text{cm}$.
        \label{fig:zoom_density2_700}}
   
\end{figure}

We finally observe that macro-scale and full scale computations agree reasonably well in the drift region, but the ionization layer near the emitter cannot be captured in the macroscopic simulations. As explained in Sec.~\ref{sec: LookUpTable} the ionization layer is vanishingly small on the scale of macroscopic computations and the boundary conditions are applied at the surface of the emitter, instead of the edge of the ionization layer. This fact is well depicted in Figs.~\ref{fig:zoom_density9_700} and~\ref{fig:zoom_density2_700}, where a zoom of the computational domain near the emitter is shown. One can clearly see the difference between macro-scale and full-scale computations. Macro-scale computations completely ignore the existence of the ionization layer and reach, at the emitter wall, a number density imposed by the injection law of Eq.~\ref{eq:fitting_exp}. On the opposite, the full-scale computations build-up progressively the number density of the positive ions inside the ionization layer and the two models agree outside the ionization region, i.e.\ in the drift region. We remark that the length of the region where the two models disagree is of the order of $10^{-4}\,\text{m}$ which is negligible with respect to the gap length, that is of the order of $10^{-2}\,\text{m}$ for the smallest gap.

\section{Two-dimensional macroscopic simulations}
\label{sec: Macroscopic2D}

\begin{figure}
\centering
    \includegraphics[width=0.65\textwidth]{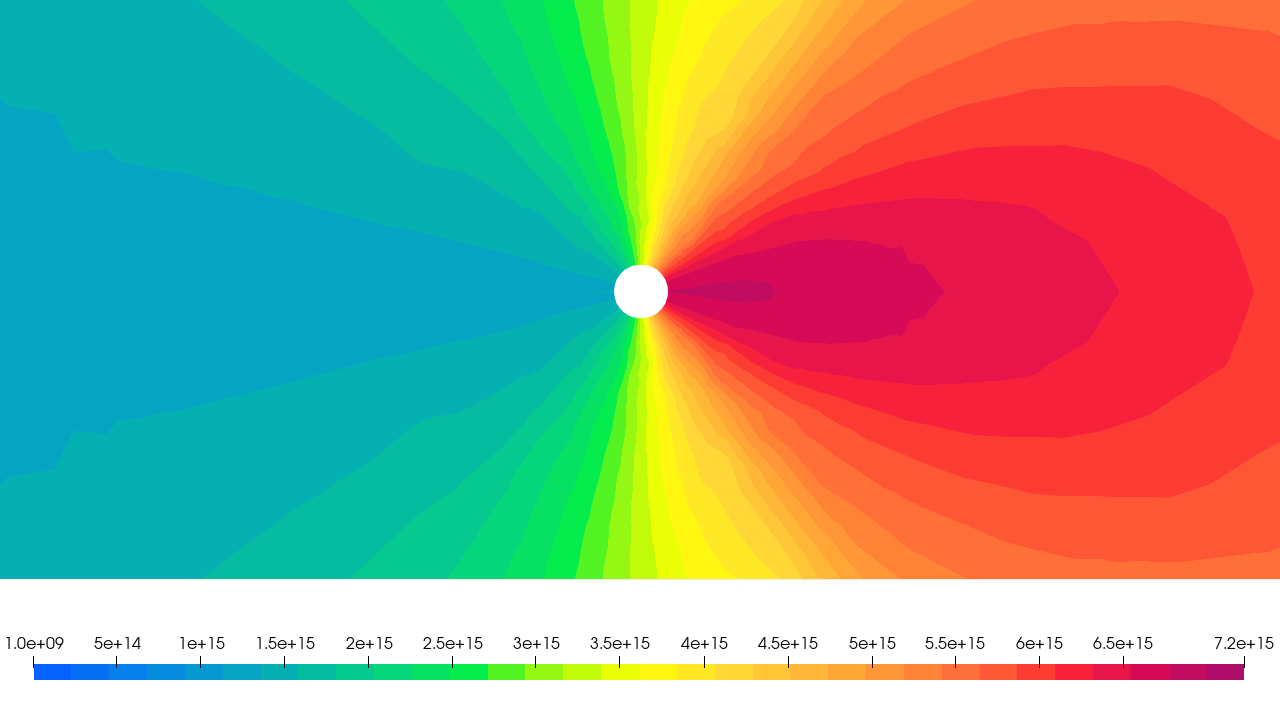}
\caption{ Number density profile of positive ions at the emitter.}
    \label{fig: zoom_emitter_ww}
\end{figure}

\begin{figure}[H]
\centering
    \includegraphics[width=0.6\textwidth]{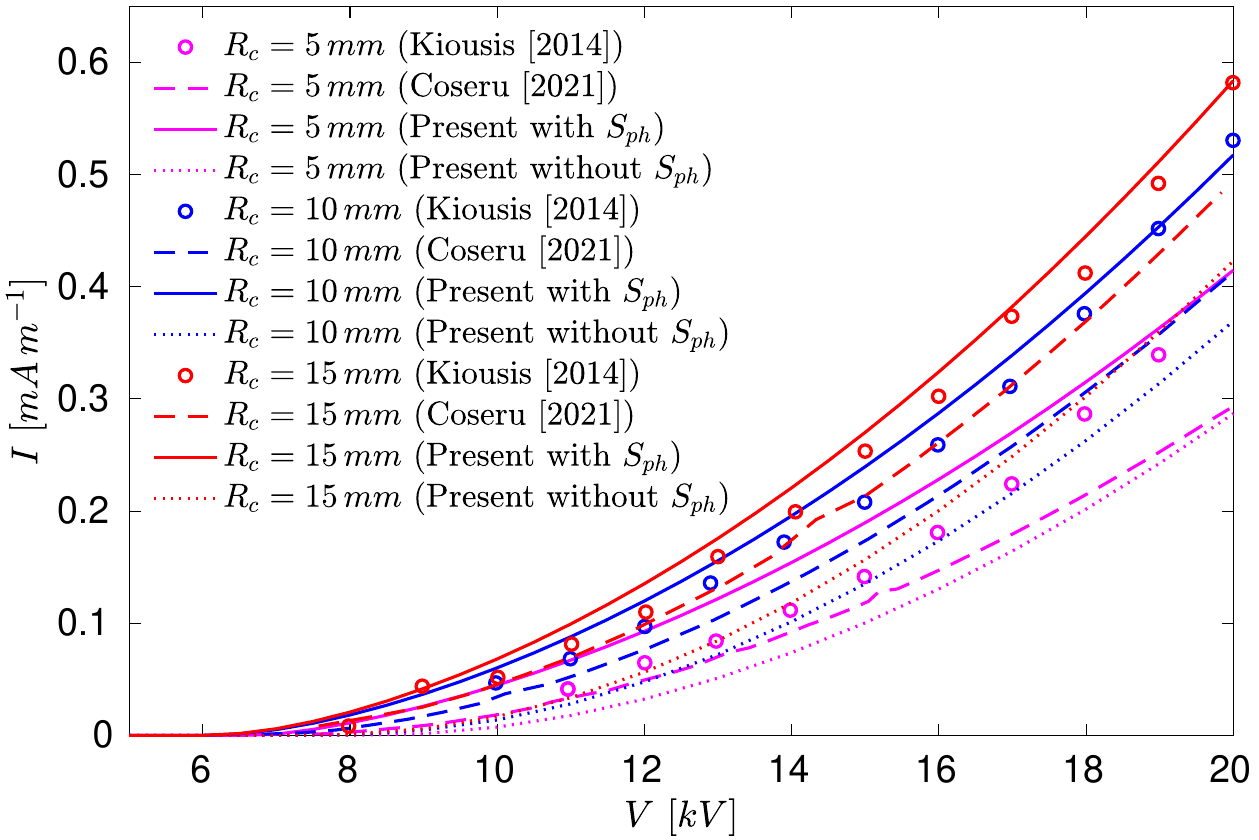}
    \caption{Current-voltage curves for wire-cylinder configuration with $R_e=50\, \mu\text{m}$, gap $L=30\,\text{mm}$ and varying collector radius.}
    \label{fig: WW_IV:D30_Rvar}
\end{figure}

\begin{figure}[H]
    \centering
    \includegraphics[width=0.6\textwidth]{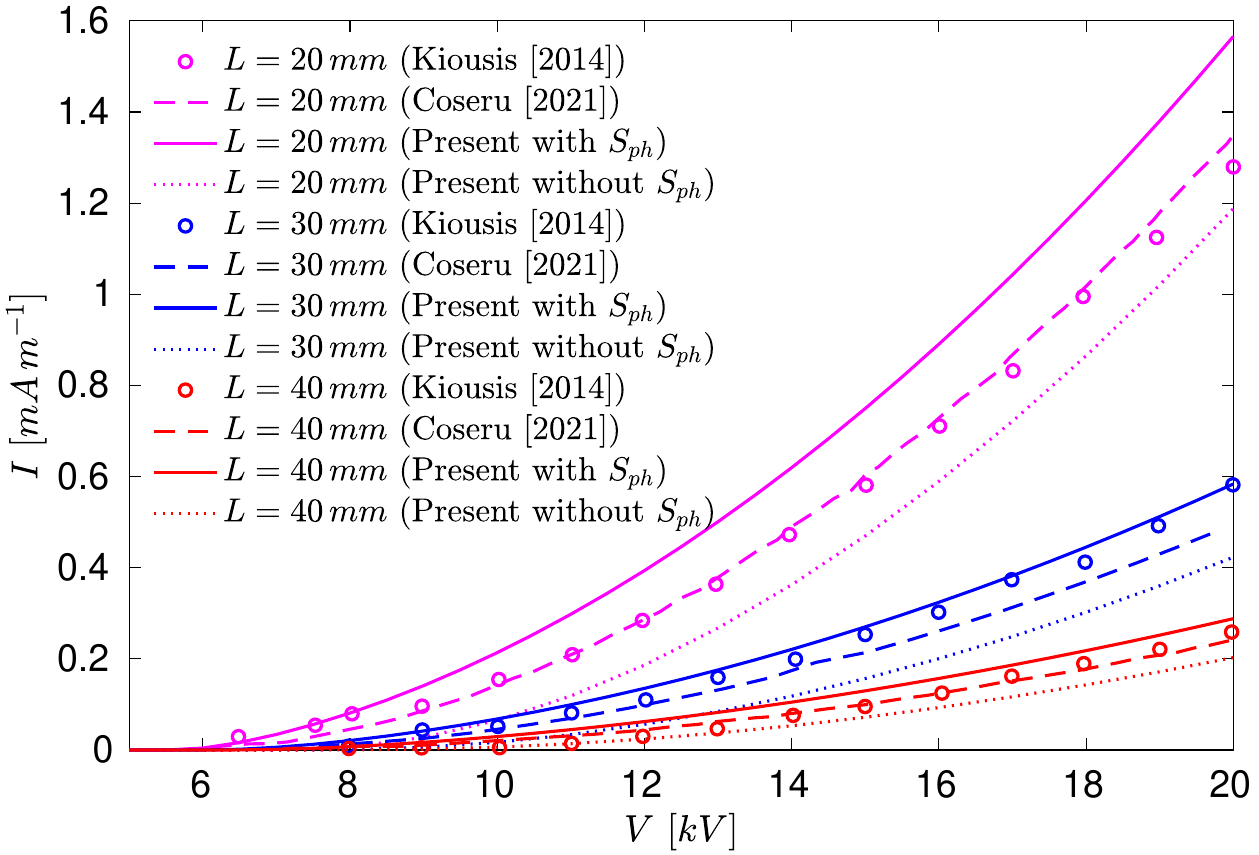}
    \caption{Current-voltage curves for  wire-cylinder configuration with $R_e=50\, \mu\text{m}$, collector $R_c=15\,\text{mm}$ and varying gap length.}
    \label{fig: WW_IV:Dvar_R15}
\end{figure}

We test the described method to compute boundary conditions for the macroscopic model in a two-dimensional wire-to-cylinder configuration, that, although not aerodynamically efficient, is similar to the configurations under study for ionic propulsion~\cite{Belan}. The two-dimensional simulations are run on an in-house code; the detailed description of the numerical methods adopted is beyond the scope of the current work.
The code uses the same modeling assumptions of Sec.~\ref{sec:macromodel} and a numerical method that, like in Sec.~\ref{sec:num_macromodel}, solves the discretized version of Eqs.~\ref{eq: DD_macro} and~\ref{eq: Poisson_macro} via Newton's method.\\
The extension to the two-dimensional setting is based on the assumption that the amount of injected charge is locally dependent
on the normal component of the electric field at the emitter surface and on the local curvature of the surface~\cite{Cagnoni}. 
A consequence of such assumptions is that the injection law can be applied point wise on the emitter, with parameters derived from one-dimensional simulations. The main purpose of the test being carried out in the present section is to assess the accuracy of simulations based on such an assumption.\\
We consider as emitter a wire of radius $R_e=50\,\mu\text{m}$, with varying gap length $L$ and collector radius $R_c$. The parameters $\Bar{E}_{on}$, $E_{ref}$ and $n_{ref}$ for such an emitter are reported in Table~\ref{tab: fitting_parameters}.\\
As an example of this kind of computations, the number density of the positive ions near the emitter is shown in Fig.~\ref{fig: zoom_emitter_ww}, for an applied voltage $V=20\, \text{kV}$.
The computations are carried out for $L=30\,\text{mm}$ and $R_c=15\,\text{mm}$.
One can notice that the density distribution is not symmetrical, because the electric field around the emitter is not symmetric too. It is remarkable that the boundary condition implemented by Eq.~\ref{eq:fitting_exp} is able to properly set the ion emission around the emitter, catching the non-symmetrical feature of such a configuration. Unlike existing literature models~\cite{Coseru}, this result is achieved without any sort of numerical regularization.\\
We compare our results with the numerical simulations of Coseru~\cite{Coseru} and the experimental results of Kiousis~\cite{Kiousis} in terms of current versus applied voltage. The results of Fig.~\ref{fig: WW_IV:D30_Rvar} are obtained fixing the gap distance to $L=30\, \text{mm}$ and varying the collector radius $R_c$ according to the values reported in the legend. Figure~\ref{fig: WW_IV:Dvar_R15} shows the curves of current versus applied voltage obtained with a collector radius $R_c=15\, \text{mm}$ and varying the gap $L$. Our results are labeled in the legend with the keywords "Present with $S_{ph}$" and "Present without $S_{ph}$", differentiating between the cases in which the boundary condition is fitted from the full scale model with and without photo-ionization respectively.

Overall our results with photo-ionization (continuous line in Fig.~\ref{fig: WW_IV:D30_Rvar} and~\ref{fig: WW_IV:Dvar_R15}) are in good agreement with the experimental data, except the case of gap length $L=20\, \text{mm}$ and collector radius $R_c=15\,\text{mm}$, where the current is overestimated by a maximum of $15\,\%$. As already remarked in Sec.~\ref{sec: Macroscopic}, when photo-ionization is neglected in the full scale model and therefore in the boundary condition for the macro-scale model, the current is under-estimated for the same applied voltage, as shown by the dotted lines of Fig.~\ref{fig: WW_IV:D30_Rvar} and~\ref{fig: WW_IV:Dvar_R15}. It is fair to conclude that these results prove the soundness of our approach.

\section{Conclusions}

In this paper, the corona discharge in a cylindrical configuration is studied using two different models. The first is a full-scale model, which includes chemical kinetics with and without photo-ionization and solves together the ionization region and the drift region. The second is a macro-scale model, which solves only the drift region and includes the ionization region via an injection law in the form of a Dirichlet boundary condition, that prescribes the density of positive ions at the end of the ionization region as a function of the electric field at the emitter surface.
The rationale of this approach is that the injection law mainly depends on the emitter geometry, as demonstrated by computations with the full-scale model. 
The relation between the positive ions number density at the end of the ionization region and the electric field is not assumed a priori, but rather computed with the full-scale model. When photo-ionization is present, the results agree quite well with the Kaptzov's hypothesis. When photo-ionization is absent, we notice deviations from the Kaptzov's hypothesis. This point is interesting and will be the object of future work. An injection law of exponential form proves to be suitable for providing both a good fit for the boundary condition data and numerical stability of the macro-scale model. Compared with the full-scale model results, the macro-scale model yields accurate results in terms of both current and positive ions number density, despite its simplicity. The exponential boundary condition is then used in a two-dimensional wire-to-cylinder configuration. Our results in terms of current and density of positive ions agree with numerical and experimental results from literature. These two-dimensional results open the way for the application of this methodological framework to more complex configurations which, coupled with the Navier-Stokes fluid equations, will be the object of future studies.
\section{Acknowledgements}
This project has received funding from the European Union’s Horizon 
Europe Research and Innovation Programme under grant agreement No. 101098900.\\

Carlo de Falco also acknowledges Financial support from the MUR  PRIN project
\textit{\``Transport phonema in low dimensional structures: models, simulations and theoretical aspects\''} CUP E53D23005900006.\\

 Carlo de Falco is a member of the Gruppo Nazionale Calcolo Scientifico-Istituto Nazionale di Alta Matematica (GNCS-INdAM).

The simulations discussed in this work were, in part, 
performed on the HPC Cluster of the Department of Mathematics of Politecnico di Milano which
was funded by MUR grant Dipartimento di Eccellenza 2023-2027.


\addtocontents{toc}{\vspace{2em}} 
\appendix
\begin{appendices}
\section{Influence of injection law parameters on the predictions of the macroscopic model}
\label{sec:appa}
\begin{figure}[H]
    \centering
    \includegraphics[width=0.6\textwidth]{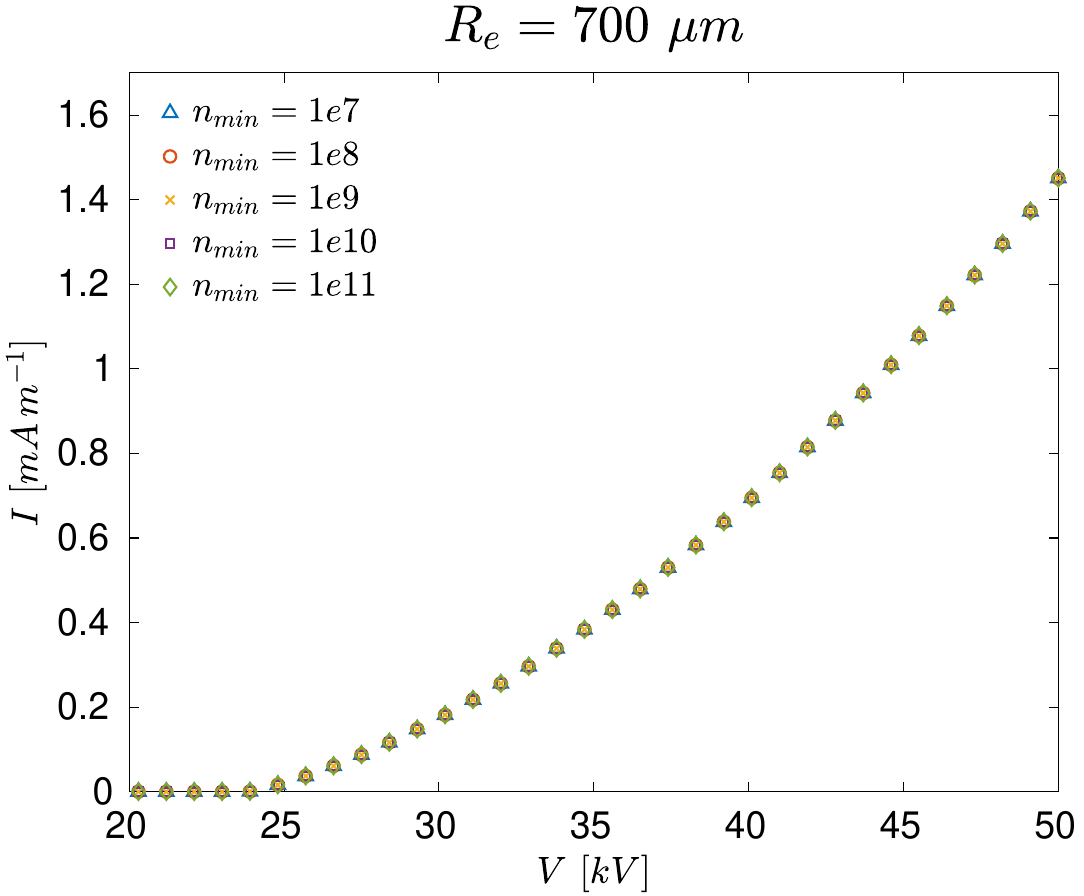}
    \caption{Current density in an axi-symmetric configuration for different values of $n_{min}$. Emitter radius $R_e=700,\mu\text{m}$; Gap $L=10.35\,\text{cm}$. Linear scale.}
    \label{fig: nmin_effect:lin}
\end{figure}

\begin{figure}[H]
\centering
    \includegraphics[width=0.6\textwidth]{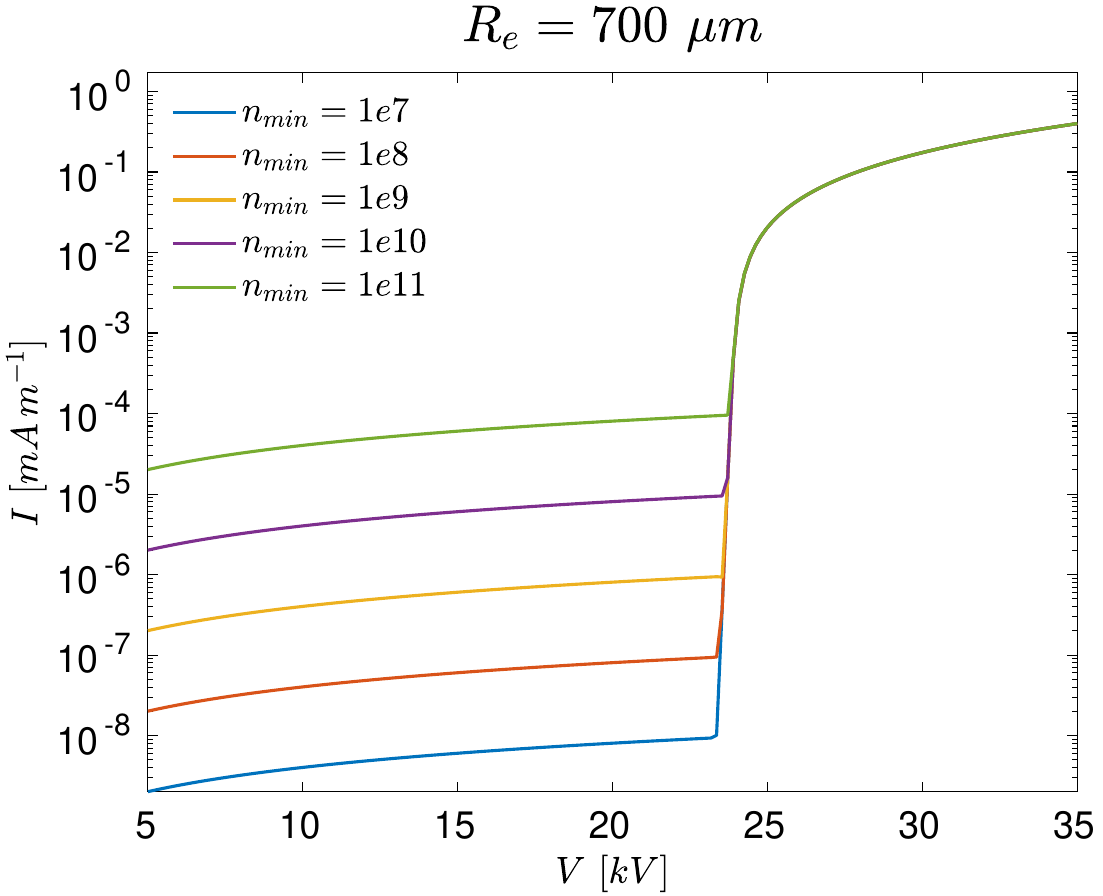}
    \caption{Current density in an axi-symmetric configuration for different values of $n_{min}$. Emitter radius $R_e=700,\mu\text{m}$; Gap $L=10.35\,\text{cm}$. Semi-log scale.}
    \label{fig: nmin_effect:log}
\end{figure}

In Section~\ref{sec: LookUpTable} we devised an injection law for positive corona discharges, based on one-dimensional full-scale simulations. The parameters $n_{ref}$, $\Bar{E}_{on}$ and $E_{ref}$ (see Eq.~\ref{eq:fitting_exp}) are chosen as a compromise between a faithful reproduction of full-scale results and the numerical stability of the macroscopic model. Another parameter, $n_{min}$, appears in the boundary conditions of the macroscopic model (see Eqs.~\ref{eq: bc_emi_macro} and \ref{eq: nmin}).
In this Appendix we aim to show how the choice of these parameters impacts the current predicted by the macroscopic model.\\
Let us first consider the effect of $n_{min}$. This parameter can be seen as a seeding of positive charges that fixes the minimum amount of ions in the computational domain. Its value has an impact only below the discharge onset and therefore it does not affect the computed current above the onset. We test the influence of the value of $n_{min}$ on the configuration of Ref.~\cite{Zheng}. We set all the parameters for the injection law equal to the values of Table~\ref{tab: fitting_parameters}, namely $\Bar{E}_{on}=6.693\times10^6$, $E_{ref}=1.3386\times 10^4$ and $n_{ref}=10^9$. We consider five different values of $n_{min}$ ($10^{11}, 10^{10}, 10^{9}, 10^8, 10^7$), as shown in Figs.~\ref{fig: nmin_effect:lin} and \ref{fig: nmin_effect:log}.
When the results are plotted in linear scale, as in Fig.~\ref{fig: nmin_effect:lin}, differences in the computed current are undistinguishable. In Fig.~\ref{fig: nmin_effect:log} we show the same results for the current, but with a logarithmic scale. We remark that $n_{min}$ does not affect the current after the onset. We recall that the onset, or ignition, voltage $V_{on}$ can be accurately computed as the solution of a suitable eigenvalue problem~\cite{benilov_onset}. In Sec.~\ref{sec:validation} we have shown that an empirical, but for our purposes acceptable, criterion is to define the onset of the discharge when the computed current reaches a threshold value set equal to $10^{-6}\, \text{A/m}$. An inspection of Fig.~\ref{fig: nmin_effect:log} shows that values of $n_{min}$ equal or greater to $10^{10}$ will result in a current in excess of $10^{-6}\, \text{A/m}$ below the onset of the discharge. The biggest value of $n_{min}$ that satisfies our current threshold is $10^9$ and this is the value we have used for every simulation performed with the macroscopic model.


\begin{figure}[H]
    \centering
    \includegraphics[width=0.6\textwidth]{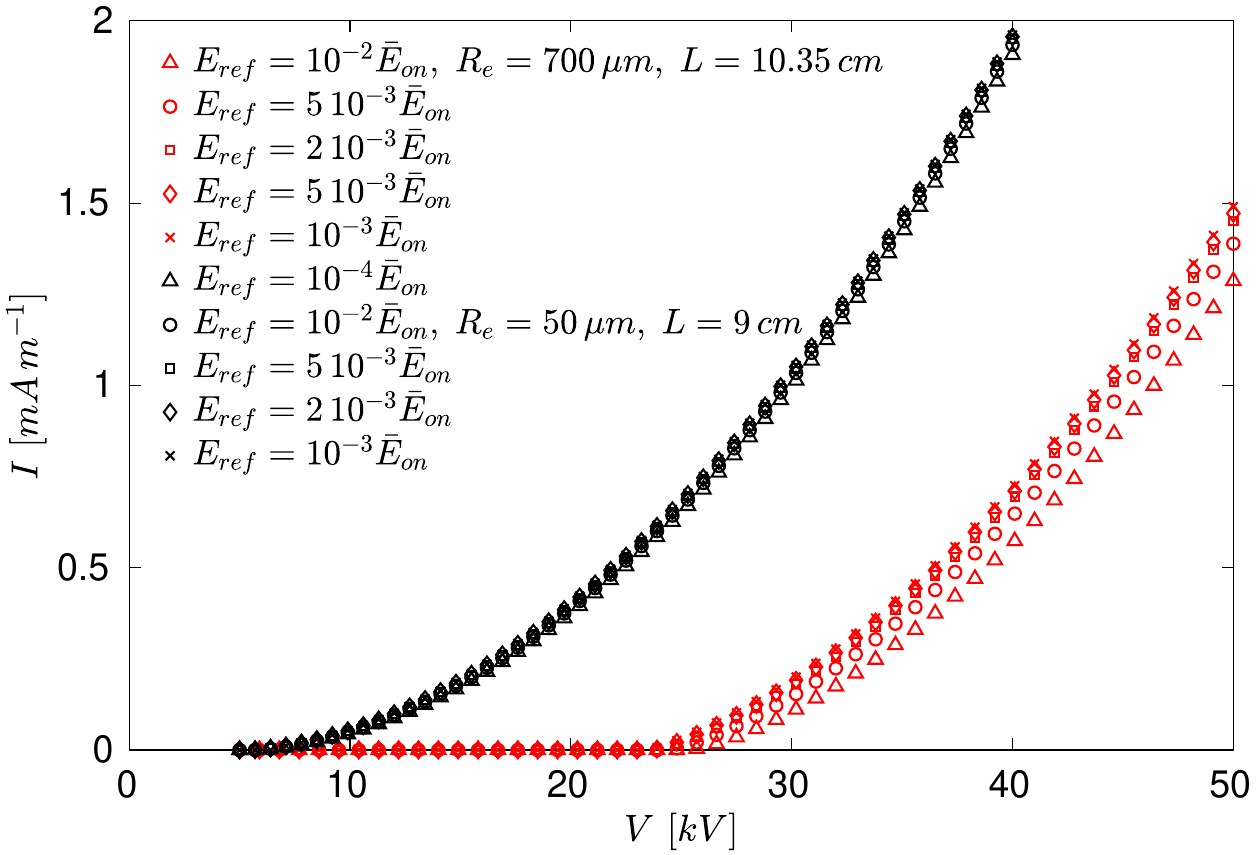}
    \caption{Effect of the choice of $E_{ref}$ on the current computed with the macroscopic model. Photo-ionization is included in the full-scale simulations.}
    \label{fig: ErefEffect}
\end{figure}

Let us now consider the effect of the choice of $E_{ref}$. As explained in Sec.~\ref{sec: LookUpTable}, it determines the slope of the functional relation between the local electric field at the emitter and the amount of charges injected. When photo-ionization is accounted for in the full-scale model, this relation turns out to be an almost vertical line (see Fig.~\ref{fig:np_vs_elec}). On the one hand it is not possible to set $E_{ref}=0$; on the other hand our goal is to relax the Kaptzov's hypothesis and to set $E_{ref}\neq 0$ to enhance the numerical stability of the macroscopic model. The right value of $E_{ref}$ is to be found empirically, as a compromise between accuracy of the computed current and numerical stability.
In Fig.~\ref{fig: ErefEffect} we show the effect of the choice of $E_{ref}$ on the computed current. We compute two different configurations, one with $R_e=50\,\mu\text{m}$ and the other with $R_e=700\,\mu\text{m}$; considering values of $E_{ref}$ that span the interval from $10^{-2}\, \Bar{E}_{on}$ up to $10^{-4}\, \Bar{E}_{on}$. We recall that $\Bar{E}_{on}$ is a fitting parameter that can be different from the true onset electric field $E_{on}$. However the two quantities are quite close for computations with photo-ionization, as it is shown by a comparison of Tables~\ref{tab: OnsetErrors} and~\ref{tab: fitting_parameters}.
The onset electric field decreases with the emitter radius (see Table~\ref{tab: OnsetErrors}). For a value of emitter radius $R_e=700\,\mu\text{m}$, setting $E_{ref}=10^{-2}\, \Bar{E}_{on}$ results in a poor prediction of the current, due to an over-relaxation of the Kaptzov's hypothesis. 
For smaller emitters, like the case $R_e=50\,\mu\text{m}$, a value of $E_{ref}=10^{-2}\, \Bar{E}_{on}$ gives already a good prediction of the current. In Table~\ref{tab: errors_Eref} we show the maximum error between the current computed with the macroscopic model and the one computed with the full-scale model, for different choices of $E_{ref}$ and two values of the emitter radius $R_e$.
\begin{table}[H]
    \footnotesize
    \centering
     \caption{Maximum error of the computed current for different values of $E_{ref}$.}
    \begin{tabular}{cccccc}
       $R_e$ [$\mu\text{m}$] & 
       $E_{ref}=10^{-2}\Bar{E}_{on}$ & 
       $E_{ref}=5\, 10^{-3}\Bar{E}_{on}$ &
       $E_{ref}=2\, 10^{-3}\Bar{E}_{on}$ & 
       $E_{ref}=10^{-3}\Bar{E}_{on}$ & 
       $E_{ref}=10^{-4}\Bar{E}_{on}$  \\ 
        \hline
        \rule{0pt}{1.5em}
        $50$    &$3.62\,\%$ &$2.24\,\%$ &   $1.41\,\%$    & $1.13\,\%$     & $0.89\,\%$\\[0.5em]
        $700$   &$14.16\,\%$  & $7.36\,\%$ &$3.21\,\%$  & $ $1.82\,\%    & $0.56\,\%$\\[0.5em]
    \end{tabular}
    \label{tab: errors_Eref}
 
\end{table}
For an emitter $R_e=700\,\mu\text{m}$, the choice of $E_{ref}=10^{-2}\, \bar{E}_{on}$ leads to an error on the computed current equal to $14.16\,\%$. On the other hand, for the smallest emitter, a value of $E_{ref}=5\times 10^{-3}\,\Bar{E}_{on}$ gives already a good agreement in the computed current, with an error equal to $2.24\,\%$. Therefore, as a compromise between accuracy and numerical stability fo the macroscopic model, we have set $E_{ref}=2 \times10^{-3}\, \Bar{E}_{on}$ for the simulations with $R_e=700\,\mu\text{m}$ and $R_e=300\,\mu\text{m}$; and $E_{ref}=10^{-2}\, \Bar{E}_{on}$ for the simulations with $R_e=50\,\mu\text{m}$.
\end{appendices}

\bibliography{bibliography} 

\end{document}